\documentclass[12pt]{iopart}

\usepackage{iopams}  
\usepackage{hyperref}
\usepackage{graphicx}
\usepackage{multirow}

\newcommand{\be}{\begin{equation}}
\newcommand{\ee}{  \end{equation}}
\newcommand{\ba}{\begin{eqnarray}}
\newcommand{\ea}{  \end{eqnarray}}
\newcommand{\ve}{\varepsilon}

\begin{document}

\title{Effective Field Theory of Emergent Symmetry Breaking in
  Deformed Atomic Nuclei}

\author{T.~Papenbrock$^{1,2}$ and H.~A. Weidenm{\"u}ller$^{3}$}

\address{$^1$Department of Physics and Astronomy, University of
  Tennessee, Knoxville, TN 37996, USA}

\address{$^2$Physics Division, Oak Ridge National Laboratory, Oak
  Ridge, TN 37831 USA}

\address{$^3$Max-Planck Institut f\"ur Kernphysik, D-69029
  Heidelberg, Germany}

\eads{\mailto{tpapenbr@utk.edu}, \mailto{haw@mpi-hd.mpg.de}}

\begin{abstract}
  Spontaneous symmetry breaking in non-relativistic quantum systems
  has previously been addressed in the framework of effective field
  theory.  Low-lying excitations are constructed from Nambu-Goldstone
  modes using symmetry arguments only. We extend that approach to
  finite systems.  The approach is very general. To be specific,
  however, we consider atomic nuclei with intrinsically deformed
  ground states. The emergent symmetry breaking in such systems
  requires the introduction of additional degrees of freedom on top of
  the Nambu-Goldstone modes. Symmetry arguments suffice to construct
  the low-lying states of the system. In deformed nuclei these are
  vibrational modes each of which serves as band head of a rotational
  band.
\end{abstract}

\maketitle

\section{Introduction}

In this paper we present a detailed approach towards emergent symmetry
breaking in finite non-relativistic quantum systems that uses concepts
of effective field theory
(EFT)~\cite{leutwyler1994,weinbergbook,brauner2010}. A short summary
of this approach was previously presented in
Ref.~\cite{papenbrock2014}, but without any details.  Spontaneous
symmetry breaking and the use of EFT in infinitely extended
non-relativistic quantum systems such as ferromagnets is well
established~\cite{roman1999,baer2004}. Our work is specifically
tailored for finite systems. Examples are finite Bose-Einstein
condensates or BCS superconductors with a broken $U(1)$ phase symmetry
and molecules and nuclei that possess deformed ground states and,
thus, break rotational invariance. In finite systems there is no
spontaneous symmetry breaking in the strict sense, and we speak
instead of emergent symmetry breaking~\cite{yannouleas2007}. The
description of such systems within an EFT requires, aside from the
standard Nambu-Goldstone modes, additional degrees of freedom. To be
specific we address here the case of deformed nuclei. The additional
degrees of freedom then describe rotations of the entire nucleus. The
treatment of emergent symmetry breaking in other systems requires only
minor modifications, however.

To set the stage we recall some aspects of the theory of nuclei with
deformed ground states such as occur in rare-earth nuclei and in the
actinides. Low-lying excitations are traditionally described
phenomenologically as rotations and vibrations, both in the collective
geometric model~\cite{bohr1975} and in the algebraic model~\cite{iachello}. Both
models describe certain ``leading order'' aspects (or gross features)
very well: Low-lying excitations are vibrational states that serve as
band heads of rotational bands. Electromagnetic intra-band transitions
are very strong, inter-band transitions are much weaker. These models
typically fail to account quantitatively for finer details regarding
``next-to-leading order'' effects such as the change of the moment of
inertia with the vibrational band head or the magnitude of weak $E2$
inter-band transitions. (Both deficiencies can be addressed within an effective theory~\cite{zhang2013,coello2015}.)

In this paper we take the view that nuclear ground-state deformation
(and, incidentally, pairing) are emergent phenomena. That view is
supported by first {\it ab initio} calculations of rotational bands in
light nuclei~\cite{caprio2013,dytrych2013}. Similarly, nuclear
mean-field calculations yield microscopic evidence for ground-state
deformations~\cite{nazarewicz1994}. For such microscopic approaches the
description of emergent phenomena is a challenge, however, because
they present multi-scale problems. Therefore, the traditional approach
within the Wigner-Weyl (linear) realization of the underlying symmetry
requires very large model spaces to describe the emergent symmetry
breaking.

The approach using an EFT is based entirely upon symmetry
arguments. The emergent broken symmetry is treated using the
nonlinear Nambu-Goldstone realization of the symmetry (as opposed to the
linear Wigner-Weyl realization) plus those additional degrees of freedom that
account for rotations. The combined treatment of these degrees of
freedom is the novel technical aspect of our work. The approach has
the advantage of being model independent. A controlled expansion in
terms of well-defined small parameters generates higher-order terms is
a systematic fashion. These may serve as useful guides in the
phenomenological models.

Aside from presenting details of the EFT approach outlined in
Ref.~\cite{papenbrock2014}, it is the purpose of this paper to show how such
higher-order terms are generated and how they influence results of
lower order. Moreover we establish the connection of our EFT with the
technically simpler effective theory of deformed nuclei developed in
Ref.~\cite{papenbrock2011}. Our results show that the effective theory is
based upon a solid field-theoretical foundation.

\section{General Approach}
\label{gen}

We consider a deformed liquid drop with a space--fixed center of mass
and with axial symmetry about the body--fixed $z'$--axis. In the
present Section we introduce the Nambu-Goldstone modes
and in addition the time-dependent modes that describe rotational
motion. We warn the reader ahead of time that in Section~\ref{another}
we switch to another parameterization of these modes. The
parameterization used in the present Section is physically transparent
and easy to justify. It has the drawback that it leads to analytically
cumbersome expressions. The parameterization used in
Section~\ref{another} is easy to handle but based on a
  different physical picture. In order to exhibit the general scheme we follow the
approach of the present Section up to the construction of the
classical field theory and its symmetries. The developments of
Section~\ref{another} then run in parallel to these developments.

\subsection{Nambu-Goldstone Modes}
\label{nambu1}

We denote the Cartesian coordinates in the space--fixed system $S$
(the body--fixed system $S'$) by $\{x, y, z\}$ (by $\{x', y', z'\}$,
respectively). The nuclear ground state is invariant under $SO(2)$
rotations about the body--fixed $z'$--axis while $SO(3)$ symmetry is
broken by the deformation. The Nambu-Goldstone modes parameterize the
two-dimensional coset space $SO(3) / SO(2)$ and depend on two complex
fields $\pi_\pm$. In the body--fixed system, a mass element of the
liquid drop has spherical coordinates $(r, \theta, \phi)$. The
classical fields $\pi_\pm$ depend on these dynamical variables and on
time $t$. We neglect the $r$-dependence of $\pi_\pm$ because radial
vibrations of the liquid drop are expected to have higher excitation
energies than surface vibrations. We denote the remaining coordinates
$\theta, \phi$ and time $t$ jointly by $\rho_\mu$ with $\mu = 1, 2,
3$. The fields $\pi_\pm$ generate the local rotation
\be
U(\rho) = \exp \{ - i \pi_-(\rho) P_+ - i \pi_+(\rho) P_- \} \ .
\label{F1}
\ee
We follow Ref.~\cite{varshalovich1988} and denote the total angular momentum
operator in the body-fixed system by ${\vec P}$ and in the space-fixed
system by ${\vec J}$. Then $P_{x'}, P_{y'}, P_{z'}$ are the components
of ${\vec P}$ in the body-fixed system, and
\be
P_\pm = \mp \frac{1}{\sqrt{2}} (P_{x'} \pm i P_{y'}) \ .
\label{F2}
\ee
Although we use the expression ``angular momentum operator'' for
$\vec{P}$ and ${\vec J}$, the three components of these operators
represent in the present context only the three generators of
rotations in the body--fixed and in the space-fixed system,
respectively. With $\pi_+ = \pi^*_-$, the transformation $U$ is
unitary. For a physical interpretation of the fields $\pi_\pm$ we
write
\be
\pi_\pm = \frac{1}{2} \omega \exp \{ \pm i \zeta \}
\label{F3}
\ee
with $\omega$ and $\zeta$ as the new real dynamical fields so that
Eq.~(\ref{F1}) becomes
\be
U = \exp \{ - i \omega [ \cos \zeta P_{x'} + \sin \zeta P_{y'} ] \} \ .
\label{F4}
\ee
We interpret Eq.~(\ref{F4}) geometrically. In the body--fixed system,
rotations are around an axis perpendicular to the symmetry axis. The
azimuthal angle of the rotation axis is denoted as $\zeta$.  The angle
$\omega$ is the angle of rotation about that axis. Replacing $\zeta$
by $\zeta + \pi$ we change the orientation of the axis of rotation or,
equivalently, the direction of rotation (clockwise rotation $\to$
counter--clockwise rotation).  Clockwise rotation by the angle
$\omega$ is equivalent to counter--clockwise rotation by the angle $2
\pi - \omega$.  Therefore, we might confine the angle $\omega$ to $0
\leq \omega \leq \pi$. For reasons that will become apparent later we
do not adopt that choice here so that the ranges of the angles $\zeta$
and $\omega$ are $0 \leq \omega, \zeta \leq 2 \pi$.  We expect the
effective Lagrangian to depend on trigonometric functions of $\omega$
and $\zeta$ only. The fields $\omega$ and $\zeta$ are functions of the
dynamical variables $\rho_\mu$, $\mu = 1, 2, 3$.  Inspection shows
that a pure time dependence of $\omega$ and $\zeta$ describes the
rotation of the deformed nucleus as a whole while a genuine dependence
of $\omega$ and $\zeta$ on the angles $\theta$ and $\phi$ accounts for
surface vibrations, i.e., local dislocations of the constituents of
the deformed drop (fluid elements or nucleons, as the case may be).
Therefore, we write
\ba
\omega &=& \omega_0(t) + \omega_1(\theta, \phi, t) \ ,
\nonumber \\
\zeta &=& \zeta_0(t) + \zeta_1(\theta, \phi, t) \ ,
\label{F5}
\ea
with the understanding that $\omega_1$ and $\zeta_1$ possess a
non--trivial dependence on at least one of the angles $\theta, \phi$
(so that not all partial derivatives with respect to these angles
vanish identically). We expect (and verify later) that the functions
$\omega_1$ and $\zeta_1$ define the true Nambu--Goldstone modes of the
deformed nucleus that describe surface vibrations. For these we will
use a small--amplitude approximation. The purely time--dependent
functions $\omega_0(t)$ and $\zeta_0(t)$, on the other hand, are alien
to the standard approach to spontaneously broken symmetry. They
represent the novel element that accounts for the finite size of the
system and describe rotational motion. For these functions we cannot
use the small-amplitude approximation. This situation is technically
similar to finite-volume EFTs of quantum
chromodynamics~\cite{leutwyler1987,gasser1988} and
magnets~\cite{hasenfratz1993}. Here -- and in contrast to the genuine
finite systems we consider -- the finite volume stems from limitations
of computational resources and obscures the physics of the infinite
system one is interested in. In both cases, the purely time-dpendent
``zero mode'' exhibits large fluctuations and has to be treated
separately.

We follow Ref.~\cite{roman1999} and define the Nambu--Goldstone modes
$a^\pm_\mu$ and $a^{z'}_\mu$ of our problem by writing
\be
U^\dag (\rho) i \partial_\mu U (\rho) = a^-_\mu P_+ + a^+_\mu P_- +
a^{z'}_\mu P_{z'} \ .
\label{F6}
\ee
Here $\partial_\mu = \partial / \partial \rho_\mu$ for $\mu = 1, 2,
3$.  The coefficients $a^\pm_\mu$ and $a^{z'}_\mu$ may alternatively be
written as functions of the fields $\pi_\pm$ or of the fields $\omega,
\zeta$. The effective Lagrangian is obtained by forming combinations
of these coefficients that are invariant both with respect to the
group operations and with respect to space rotations.

We express the Nambu--Goldstone modes in terms of $\omega$ and
$\zeta$. The calculation is lengthy but straightforward. We use the
left-hand side of Eq.~(\ref{F6}), the form~(\ref{F4}) for $U$, and a
Taylor expansion for $U$. We find
\ba
a^-_\mu &=& (1/2) \exp \{ - i \zeta \} [ \partial_\mu \omega - i
(\partial_\mu \zeta) \ \sin \omega ] \ ,
\nonumber \\
a^+_\mu &=& (1/2) \exp \{ + i \zeta \} [ \partial_\mu \omega + i
(\partial_\mu \zeta) \ \sin \omega ] \ ,
\nonumber \\
a^{z'}_\mu &=& (\partial_\mu \zeta) (1 - \cos \omega) \ .
\label{F18}
\ea

\subsection{Invariants}
\label{inv1}

To determine the behavior of $a^\pm_\mu$ and $a^{z'}_\mu$ under group
operations we consider the action of a fixed element $g$ (independent
of the $\rho_\mu$) of the coset space on $U$. Equation~(\ref{F1}) implies
that under the action of $g$, $U$ changes nonlinearly,
\be
U \to [g U] h^\dag(g, U) \ .
\label{F7}
\ee
The product $[g U]$ lies within the coset space $SO(3) / SO(2)$ while
\be
h(g, U) = \exp \{ i \Psi(g, U) P_{z'} \}
\label{F8}
\ee
is an element of $SO(2)$. The function $\Psi$ depends on the
parameters characterizing both $g$ and the transformation $U$. Thus,
\be
U^\dag (\rho) i \partial_\mu U (\rho) \to h U^\dag (\rho) i
[\partial_\mu U (\rho)] h^\dag + \partial_\mu \Psi(g, U) P_{z'}
\label{F9}
\ee
and
\be
a^-_\mu P_+ + a^+_\mu  P_- + a^{z'}_\mu P_{z'} \to h (a^-_\mu P_+ +
a^+_\mu  P_- + a^{z'}_\mu P_{z'}) h^\dag + \partial_\mu \Psi(g, U)
P_{z'} \ .
\label{F10}
\ee
We recall the commutation relations
\ba
&& [ J_x, J_y ] = i J_z \ ({\rm cyclic}) \ , \nonumber \\
&& [ P_{x'}, P_{y'} ] = - i P_{z'} \ ({\rm cyclic}) \ , \nonumber \\
&& [ J_k , P_{l'} ] = 0 \ ({\rm all} \ k, l = x, y, z ) \ .
\label{F11}
\ea
which imply
\ba
&& [J_z, J_+] = J_+ \ , \ [ J_z, J_- ] = - J_- \ , \ [ J_+, J_- ] =
- J_z \ , \nonumber \\
&& [P_{z'}, P_{+}] = - P_{+} \ , \ [ P_{z'}, P_- ] = P_- \ , \
[ P_+, P_- ] = P_{z'}
\label{F12}
\ea
and $P^m_{z'} P_\pm = P_\pm (P_{z'} \mp 1)^m$. To calculate the
right-hand side of expression~(\ref{F10}) we use these for $\vec{ P}$
and the expansion of $h$ in a Taylor series in $P_{z'}$. We obtain
\ba
a^-_\mu &\to& \exp \{ - i \Psi \} a^-_\mu \ , \nonumber \\ a^+_\mu
&\to& \exp \{ i \Psi \} a^+_\mu \ , \nonumber \\ a^{z'}_\mu &\to&
a^{z'}_\mu + \partial_\mu \Psi(g, U) \ .
\label{F13}
\ea
Therefore, the terms of lowest order in the $a^\pm_\mu, a^{z'}_\mu$
that are invariant under group operations are~\cite{roman1999}
\be
a^+_\mu a^-_\nu \ {\rm and} \ \partial_\mu a^{z'}_\nu - \partial_\nu
a^{z'}_\mu \ .
\label{F14}
\ee
Here $\mu, \nu = 1, 2, 3$. The expressions~(\ref{F14}) are obtained by
treating the derivatives $\partial_\mu$ both with respect to the
angles $\theta, \phi$ and with respect to time as though they were
derivatives with respect to external variables. While that is
appropriate for time $t$ ($\mu = 3$), it is not for the angles
$\theta, \phi$ ($\mu = 1, 2$) because these change under
rotations. The effective Lagrangian must be invariant with respect to
such rotations. In actually constructing the invariants we demand
axial symmetry (see Section~\ref{another}) and are guided by the
analogy to coordinate transformations in Euclidean space in three
dimensions. Here we would interpret $\partial_k$ with $k = x, y, z$ as
one component of the vector $\vec{\nabla}$, use the invariance of the
scalar product of two vectors, and obtain the invariants
$\vec{\nabla}a^+ \vec{\nabla}a^-$. On the unit sphere we analogously
replace $\vec{\nabla}$ by the vector $\vec{L}$ of orbital angular
momentum with components
\ba
L_{x'} &=& + i \sin \phi \frac{\partial}{\partial \theta} + i \cos \phi
\cot \theta \frac{\partial}{\partial \phi} \ , \nonumber \\
L_{y'} &=& - i \cos \phi \frac{\partial}{\partial \theta} + i \sin \phi
\cot \theta \frac{\partial}{\partial \phi} \ , \nonumber \\
L_{z'} &=& - i \frac{\partial}{\partial \phi} \ .
\label{F15}
\ea
The lowest--order axially symmetric invariants formed from $a^\pm$ are
then
\ba
&L_{x'} a^+ L_{x'} a^- + L_{y'} a^+ L_{y'} a^- \ , \nonumber\\
&L_{z'} a^+ L_{z'} a^- \ , \nonumber\\
&\partial_t a^+ \partial_t a^- \ .
\label{F16}
\ea
Expressions like $L_{x'} a^+$ are here understood as linear
combinations of terms $a^+_\mu$ (read as partial derivatives with
respect to $\rho_\mu$ and explicitly given in Eqs.~(\ref{F18})) with
coefficients given in the first of Eqs.~(\ref{F15}). For simplicity,
we employ in the present Section only the spherical invariant
\be
L_{x'} a^+ L_{x'} a^- + L_{y'} a^+ L_{y'} a^- + L_{z'} a^+ L_{z'} a^- \ ,
\label{F17}
\ee
and use the more general axially symmetric invariants in
Section~\ref{another}.

The substitution rules~(\ref{F13}) apply in both the body-fixed and
the space-fixed coordinate system, albeit with different definitions
of the function $\Psi$. Using this fact and the well-known behavior of
the components of $\vec{L}$ under rotations we easily confirm that the
form~(\ref{F17}) is indeed rotationally invariant.

The eigenstates of the nuclear Hamiltonian are (almost exactly)
eigenstates of the parity operator. But that operator cannot be
written in terms of rotations. Indeed, in three dimensions the
determinant of the matrix describing parity inversion equals minus one
while the determinants of all rotation matrices are equal to plus one
because these connect continuously to the unit matrix. Therefore, we
cannot incorporate parity conservation into the present approach. We
return to that point in Section~\ref{disc}. The situation for time
reversal differs. To be invariant under time reversal the effective
Lagrangian must contain only even derivatives with respect to
time. That constraint has been taken into account in
expressions~(\ref{F16}). The constraint is appropriate for even--even
nuclei with ground-state spin zero. Another discrete symmetry (the
${\cal R}$-parity of Ref.~\cite{bohr1975}) plays a role for the quantized
version of the theory and is treated in Section~\ref{disc}, too.

The explicit form of the invariants is obtained by using
Eqs.~(\ref{F18}) in expressions~(\ref{F16}) and (\ref{F17}),
\ba
{\cal{L}}_1 &=& 2 a^{-}_t a^{+}_t = \frac{1}{2} \bigg(
\dot{\omega}^2 + \dot{\zeta}^2 \sin^2 \omega \bigg) \ , \nonumber \\
{\cal {L}}_2 &=& 2 \sum_{k = 1}^3 (L_k a^+) (L_k a^-) = \frac{1}{2}
\bigg( ( \vec{L} \omega )^2 + \sin^2 \omega \ ( \vec{L} \zeta )^2
\bigg) \ .
\label{F20}
\ea
The dot indicates the time derivative.

\subsection{Classical Field Theory}
\label{class1}

The classical field theory is obtained by writing the effective
Lagrangian density ${\cal L}$ as a linear combination of the two
invariants~(\ref{F20}),
\be
{\cal{L}} = C {\cal{L}}_1 + D {\cal{L}}_2 \ .
\label{F19}
\ee
The coefficients $C$ and $D$ are the parameters of the theory and are
determined by a fit to the data. The total Lagrangian $L$ is obtained
by integrating ${\cal L}$ over the dynamical angles $\theta, \phi$,
\be
L = \int {\rm d} E \ {\cal L} \equiv \frac{1}{4 \pi}  \int_0^\pi
{\rm d} \theta \sin \theta \int_0^{2 \pi} {\rm d} \phi \ {\cal L} \ .
\label{F21}
\ee
The normalization is chosen such that for ${\cal L} = 1$ the integral
in Eq.~(\ref{F21}) yields unity. The factor $\sin \theta$ in the
measure~(\ref{F21}) takes account of the fact that $\theta, \phi$
define points on the surface of a sphere so that we deal with
curvilinear coordinates. We show in Appendix 1 that in such
coordinates the equations of motion are
\ba
\sum_\mu \partial_\mu \bigg( \frac{\partial (\sin \theta {\cal
L})}{\partial (\partial_\mu \omega)} \bigg) = \frac{\partial (\sin
\theta {\cal L})} {\partial \omega} \ , \nonumber \\
\sum_\mu \partial_\mu \bigg( \frac{\partial (\sin \theta {\cal
L})}{\partial (\partial_\mu \zeta)} \bigg) = \frac{\partial (\sin
\theta {\cal L})} {\partial \zeta} \ .
\label{F22}
\ea
We recall that $\mu = 1, 2$ stands for the angles $\theta, \phi$ and
$\mu = 3$ for the time $t$, and that both $\omega$ and $\zeta$ are
functions of $\theta, \phi, t$. Eqs.~(\ref{F22}), (\ref{F19}), and
(\ref{F20}) constitute the non--linear equations of motion for the two
classical fields $\omega$ and $\zeta$.

In constructing ${\cal L}$ we have imposed rotational invariance.
Therefore, we expect that the total angular momentum of the system is
conserved. Using the Noether theorem as in Ref.~\cite{papenbrock2011}, we now
construct three constants of the motion. These correspond to the three
components of angular momentum (not to be confused with the three
components of the rotation operator in Eqs.~(\ref{F11})). With $k =
x', y', z'$ we consider the rotation 
\be
r = \exp \{ - i \sum_k \delta \chi_k P_k \}
\label{F23}
\ee
by infinitesimally small angles $\delta \chi_k$ about the three
body--fixed axes. The ensuing infinitesimal changes of $\omega$ and of
$\zeta$ are denoted by $\delta \omega$ and $\delta \zeta$, those of
${\cal L}$ by $\delta {\cal L}$. A straightforward calculation yields
\be
\delta {\cal L} = \sum_{k = 1}^3 \delta \chi_k \bigg\{ \sum_{\mu =
1}^3 \frac{\partial}{\partial \mu} \bigg( \frac{\partial {\cal L}}
{\partial (\partial_\mu \omega)} M_{1 k} + \frac{\partial {\cal L}}
{\partial (\partial_\mu \zeta)} M_{2 k} \bigg) \bigg\} \ .
\label{F24}
\ee
With $\delta \omega = \delta q_1$ and $\delta \zeta = \delta q_2$ the
matrix $M$ is defined by $\delta q_\nu = \sum_k M_{\nu k} \delta
\chi_k$ and explicitly given in Appendix~2. Rotational invariance
implies $\delta {\cal L} = 0$ for every choice of $\delta \chi_k$ with
$k = x', y', z'$. That implies the vanishing of each of the three
Noether currents $k = x', y', z'$ in big curly brackets in
Eq.~(\ref{F24}).

To obtain the constants of the motion we consider an infinitesimal
change of the total Lagrangian given by
\be
\delta L = \int {{\rm d} \theta {\rm d} \phi \over 4 \pi} \delta 
(\sin \theta {\cal L}) \ ,
\label{F25}
\ee
with $\delta {\cal L}$ defined in Eq.~(\ref{F24}). The effective
Lagrangian density ${\cal L}$ is periodic in the angle $\phi$. With
respect to $\theta$, the same statement holds for ${\cal L} \sin
\theta$. Partial integration then shows that the derivative terms with
respect to $\theta$ and $\phi$ in Eq.~(\ref{F24}) do not contribute to
$\delta L$ so that $\delta L$ is given by a pure time derivative. The
vanishing of $\delta L$ for any choice of the angles $\delta \chi_k$
then implies the existence of three constants of the motion. These are
the three components $Q_k$ of the total angular momentum of the
system. With $M$ given in Eq.~(\ref{B14}) of Appendix~2 and the
integration measure defined in Eq.~(\ref{F21}) these are
\ba
Q_{x'} &=& \int {\rm d} E \ \bigg( \frac{\partial {\cal L}} {\partial
\dot{\omega}} \cos \zeta - \frac{\partial {\cal L}} {\partial
\dot{\zeta}} \sin \zeta \cot \omega \bigg) \ , \nonumber \\
Q_{y'} &=& \int {\rm d} E \ \bigg( \frac{\partial {\cal L}} {\partial
\dot{\omega}} \sin \zeta + \frac{\partial {\cal L}} {\partial
\dot{\zeta}} \cos \zeta \cot \omega \bigg) \ , \nonumber \\
Q_{z'} &=& - \int {\rm d} E \ \frac{\partial {\cal L}} {\partial
\dot{\zeta}} \ . 
\label{F26}
\ea

\subsection{Power Counting}
\label{pow1}

Without much justification given, we have constructed the effective
Lagrangian in Eqs.~(\ref{F19}) and (\ref{F20}) from the lowest-order
invariants in the Nambu-Goldstone modes. Further progress hinges on
the identification of the energy scales that govern our problem and of
the associated powers of higher-order terms in the Nambu-Goldstone
modes. The relevant scales are the energy scale $\xi$ of rotational
motion, the energy scale $\Omega$ of vibrational motion, and the
cutoff parameter $\Lambda$ beyond which other modes like fermionic
excitations play a role. In our approach it is assumed that $\xi \ll
\Omega \ll \Lambda$.

To identify the various contributions to the effective Lagrangian in
Eq.~(\ref{F19}) and to the equations of motion~(\ref{F22}), we use the
decomposition~(\ref{F5}) and an expansion of $\omega_1$ and of
$\zeta_1$ in terms of spherical harmonics,
\ba
\omega_1 &=& \sum_{L = 2}^\infty \sum_\mu \omega_{L \mu}(t)
Y_{L \mu}(\theta, \phi) \ , \nonumber \\
\zeta_1 &=& \sum_{L = 2}^\infty \sum_\mu \zeta_{L \mu} (t)
Y_{L \mu}(\theta, \phi) \ .
\label{F27}
\ea
The term with $L = 1$ describes center-of-mass motion and is
suppressed. As mentioned earlier we take rotational motion fully into
account by treating $\omega_0$ and $\zeta_0$ without any
approximation. We use a small--amplitude approximation for the
surface vibrations described by $\omega_1$ and $\zeta_1$. A naive
first approach would then consist in expanding ${\cal L}$ in powers of
$\omega_1$ and of $\zeta_1$ and in keeping only terms up to second
order (but terms of all orders in $\omega_0$ and in $\zeta_0$). That
would be in line with the treatment of ferromagnets and paramagnets in
Ref.~\cite{roman1999} where in the leading--order effective Lagrangian
only terms up to second order in the fields are kept. However, because
of the presence of rotational motion that procedure does not fully
apply in our case. To see that we must analyze the contributions to
${\cal L}$ in some detail. We use the symbol $\sim$ to define the
relevant order, and we drop the magnetic quantum numbers on $\omega_L$
and $\zeta_L$.

The ratios $\dot{\omega}_0 / \omega_0 \sim \xi$ and $\dot{\zeta}_0 /
\zeta_0 \sim \xi$ are governed by the energy scale of rotational
motion. The range of the variables $\omega_0$ and $\zeta_0$ is of
order unity. Therefore, $\dot{\omega}_0 \sim \xi$ and $\dot{\zeta}_0
\sim \xi$. Inserting Eqs.~(\ref{F5}) into Eqs.~(\ref{F22}) we obtain
the terms $(C / 2) \dot{\omega}^2_0$ and $(C / 2) \dot{\zeta}^2_0$.
We show below that these describe rotational motion. Therefore, $(C /
2) \dot{\omega}^2_0 \sim \xi$ and $(C / 2) \dot{\zeta}^2_0 \sim
\xi$. Together with $\dot{\omega}_0 \sim \xi$ and $\dot{\zeta}_0 \sim
\xi$ that implies $C \sim 1 / \xi$ which is consistent with the
interpretation of $C$ as moment of inertia. The ratios $\dot{\omega_L}
/ \omega_L \sim \Omega$ and $\dot{\zeta}_L / \zeta_L \sim \Omega$ with
$L \geq 2$ are similarly governed by the energy scale $\Omega$ of
vibrational motion. Inserting Eqs.~(\ref{F5}) into Eqs.~(\ref{F22}) we
obtain for $L \geq 2$ the terms $(C / 2) \dot{\omega}^2_L$ and $(C /
2) \dot{\zeta}^2_L$. These describe vibrational motion. Therefore $(C
/ 2) \dot{\omega}^2_L \sim \Omega$ and $(C / 2) \dot{\zeta}^2_L \sim
\Omega$. Together with $C \sim 1 / \xi$ these relations imply
$\dot{\omega}_L, \dot{\zeta}_L \sim \sqrt{\xi \Omega}$ and, together
with $\dot{\omega}_L / \omega_L \sim \Omega$, $\dot{\zeta}_L / \zeta_L
\sim \Omega$ also $\omega_L \sim \sqrt{\xi / \Omega}$, $\zeta_L \sim
\sqrt{\xi / \Omega}$ for $L \geq 2$.  These relations are used below
when we expand ${\cal L}$ in powers of $\omega_L$ and $\zeta_L$. The
terms in ${\cal{L}}_2$ (see Eqs.~(\ref{F20})) describe vibrational
motion so we have $(D / 2)$ $(\vec{L} \omega_L)^2 \sim \Omega$ and $(D
/ 2) (\vec{L} \zeta_L)^2 \sim \Omega$.

In our approach the operator $\vec{L}$ is dimensionless (see
Eqs.~(\ref{F15})). Formally, however, ${\vec L}$ plays the same role
as the momentum in the theory of Ref.~\cite{roman1999} where only small
momenta are kept for small energies. Physically we analogously expect
that only the small eigenvalues of the operator $\vec{L}^2$ are
relevant for the low--energy part of the spectrum in our case.
Therefore we formally attach to $\vec{L}$ the scale $\Omega$.
Together with $\omega_L, \zeta_L \sim \sqrt{\xi / \Omega}$ that gives
$D \sim 1 / \xi$. We summarize these assumptions and results by
writing
\ba
\xi &\ll& \Omega \ , \nonumber \\
\omega_0, \zeta_0 &\sim& 1 \ , \nonumber \\
\omega_L, \zeta_L &\sim& \sqrt{\xi / \Omega} \ \ll 1 \ {\rm for} \ L
\geq 2 \ , \nonumber \\
\dot{\omega_0}, \dot{\zeta}_0 &\sim& \xi \ , \nonumber \\
\ddot{\omega_0}, \ddot{\zeta}_0 &\sim& \xi^2 \ , \nonumber \\
\dot{\omega_L}, \dot{\zeta}_L &\sim& \sqrt{\xi \Omega} \ {\rm for} \ L
\geq 2 \ , \nonumber \\
\ddot{\omega_L}, \ddot{\zeta}_L &\sim& \Omega \sqrt{\xi \Omega} \
{\rm for} \ L \geq 2 \ , \nonumber \\
C &\sim& \frac{1}{\xi} \ , \nonumber \\
D &\sim& \frac{1}{\xi} \ , \nonumber \\
\vec{L} &\sim& \Omega \ .
\label{F28}
\ea
These considerations imply the following rule for an approximate
treatment of the problem. The effective Lagrangian $\cal{L}$ in
Eqs.~(\ref{F19}, \ref{F20}) contains terms of order $\Omega$ that
describe nuclear surface vibrations and terms of order $\xi$ that
describe rotational motion. In expanding $\cal{L}$ in powers of
$\omega_L$ and $\zeta_L$ we must, therefore, keep terms of orders
$\Omega$, $\sqrt{\Omega \xi}$ and $\xi$. We omit terms of order
$\sqrt{\xi / \Omega}$ or less. The resulting approximate expressions
are
\ba
C {\cal{L}}_1 &\approx& \frac{C}{2} \bigg( (\dot{\omega}_0 +
\dot{\omega}_1)^2 + \dot{\zeta}^2_0 \sin^2 \omega_0 \nonumber \\
&& + 2 \dot{\zeta}_0 \dot{\zeta}_1 \bigg[ \sin^2 \omega_0 + \omega_1
\sin 2 \omega_0 \bigg] \nonumber \\
&& + \dot{\zeta}^2_1 \bigg[ \sin^2 \omega_0 + \omega_1 \sin 2 \omega_0
+ \omega^2_1 \cos 2 \omega_0 \bigg] \bigg) \ , \nonumber \\
D {\cal {L}}_2 &\approx& \frac{D}{2} \bigg( ( \vec{L} \omega )^2 +
( \vec{L} \zeta )^2 \bigg[ \sin^2 \omega_0 + \omega_1 \sin 2 \omega_0 
\nonumber \\
&& \qquad \qquad + \omega^2_1 \cos 2 \omega_0 \bigg] \bigg) \ .
\label{F29}
\ea
These developments show which terms to keep in the effective
Lagrangian.

\subsection{Purely Rotational Motion}
\label{prm} 

To ascertain consistency of our arguments we consider the case of
nuclear rotation without surface vibrations. We accordingly assume
that $\omega$ and $\zeta$ depend only upon time, so that in
Eqs.~(\ref{F5}) we have $\omega_1 = 0 = \zeta_1$. For purely
time--dependent fields classical field theory changes into classical
mechanics. The treatment becomes very similar to that of
Ref.~\cite{papenbrock2011}. The effective Lagrangian is
\be
{\cal L}  = \frac{C}{2} \bigg( \dot{\omega}^2_0
+ \dot{\zeta}^2_0 \sin^2 \omega_0 \bigg) \ ,
\label{F30}
\ee
and the canonical momenta are
\ba
\pi_\omega &=& \frac{\partial {\cal L}} {\partial \dot{\omega}_0} =
C \dot{\omega}_0 \ , \nonumber \\
\pi_\zeta &=& \frac{\partial {\cal L}} {\partial \dot{\zeta}_0} =
C \dot{\zeta}_0 \sin^2 \omega_0 \ .
\label{F31}
\ea
The Hamiltonian $H$ is given by
\be
H = \frac{1}{2 C} \bigg( \pi^2_\omega + \frac{1}{\sin^2 \omega_0} 
\pi^2_\zeta \bigg) \ .
\label{F32}
\ee
The three components $Q_k$ of angular momentum are given by
Eqs.~(\ref{F26}) which now read
\be
Q_k = \pi_\omega M_{1 k} + \pi_\zeta M_{2 k} \ .
\label{F33}
\ee
We use Eq.~(\ref{B14}) of Appendix~2 for $M$ and find
\be
\sum_k Q^2_k = \pi^2_\omega + \frac{1}{\sin^2 \omega_0} \ \pi^2_\zeta
\ .
\label{F34}
\ee
That shows that the Hamiltonian~(\ref{F32}) is proportional to the
square of the total angular momentum,
\be
H = \frac{1}{2 C} \sum_k Q^2_k \ .
\label{F35}
\ee
In other words, we obtain the classical theory of the rotating top.
The constant $C$ is the moment of inertia.

\section{Another Parameterization}
\label{another}

We have actually carried the approach of Section~\ref{gen} further,
deriving the Hamiltonian and quantizing it. The resulting equations
are difficult to interpret, however. They do not display in an obvious
fashion what is expected on physical grounds: Harmonic vibrational
motion of the variables $\omega_L$ and $\zeta_L$. As shown in
Appendix~3, these difficulties have to do with the non--Cartesian form
of the measure ${\rm d} E$ in Eq.~(\ref{F21}). That is why we now
introduce another parameterization of the matrix $U$ defined in
Eq.~(\ref{F1}). We proceed in close analogy to Section~\ref{gen}.

\subsection{Nambu-Goldstone Modes}
\label{nambu2}

We use the space-fixed system, and we parameterize the matrix $U$ in
product form,
\ba
U &=& g(\zeta, \omega) \ u(x,y) \ , \nonumber\\
g(\zeta, \omega) &=& \exp \left\{ -i \zeta(t) \hat{J}_z \right\}
\exp\left\{ -i \omega(t) \hat{J}_y \right\} \ , \nonumber\\
u(x,y) &=& \exp\left\{-i x \hat{J}_x
-i y \hat{J}_y \right\} \ .
\label{F36}
\ea
The purely time--dependent variables $\omega$ and $\zeta$ describe
rotations of the finite system, similarly to the variables $\omega_0,
\zeta_0$ introduced in Eqs.~(\ref{F5}). As in Section~\ref{gen} we
choose the ranges as $0 \leq \omega, \zeta \leq 2 \pi$. This is
convenient for Section~\ref{quant}. With $\theta$ and $\phi$ as
defined in Section~\ref{gen}, the fields $x = x(\theta, \phi, t)$ and
$y = y(\theta, \phi, t)$ play the role of the fields $\omega_1$ and
$\zeta_1$ defined in Eqs.~(\ref{F5}). They describe the
small-amplitude vibrations of the liquid drop. To exclude the
possibility that $x$ and $y$ induce a global rotation of the entire
drop we request
\be
\int {\rm d} E \ x(\theta, \phi, t) = 0 = \int {\rm d} E \
y(\theta, \phi, t) \ .
\label{F37}
\ee
We use the definition that $U = g u$ acts onto objects to the
right. Thus, the local vibrations induced by the field $u$ are
followed by a global rotation $g$ of the entire drop. 

We show in Appendix 3 that the parameterization of $U$ in terms of the
variables $\omega$ and $\zeta$ in Eqs.~(\ref{F4}) and (\ref{F5}) and
the one introduced in Eq.~(\ref{F36}) are completely equivalent. Why
then did we not start from the outset with the
parameterization~(\ref{F36})? As shown in Section~\ref{gen}, the
parameterization used in Eqs.~(\ref{F4}) and (\ref{F5}) can be
justified physically in a convincing manner. Moreover, it is tailored
after the standard approach to symmetry breaking in non-relativistic
systems. The advantage of the new parameterization is that it treats
the rotational degrees of freedom separately while the
parameterization in Eqs.~(\ref{F4}) and (\ref{F5}) treats the
rotational mode and the vibrational modes on an equal footing.  It has
an alternative physical interpretation. When acting from right to
left, $u$ induces a small-amplitude dislocation of a nucleon (or
volume element) at $(\theta,\phi)$ in the axially-symmetric nucleus
(whose symmetry axis is the $z$ axis), while $g$ then rotates the
entire nucleus.

Power counting as in Section~\ref{pow1} shows that
\ba
\omega, \zeta &\sim& {\cal O}(1) \ , \nonumber \\
\dot\omega, \dot \zeta &\sim& \xi \ , \nonumber \\
|x|, |y| &\sim& \ve^{1/2} \ll 1 \ , \nonumber \\ 
\dot{x}, \dot{y} &\sim& \Omega \ve^{1/2} \ .
\label{F38}
\ea
The parameter $\ve$ helps to identify (and omit) higher powers of $x$
and $y$, consistent with a focus on small-amplitude harmonic surface
vibrations. A physical interpretation of this parameter is given at
the end of Section~\ref{class2} below. From here on the development is
similar to that of Section~\ref{gen}.

In analogy to Eq.~(\ref{F6}) we define
\ba
U^{-1} i \partial_\mu U &=& a_\mu^x J_x + a_\mu^y J_y + a_\mu^z J_z \ .
\label{F39}
\ea
As in Section~\ref{gen} the symbol $\partial_\mu$ with $\mu = 1, 2, 3$
stands for the partial derivatives with respect to the angles $\theta,
\phi$ and time $t$ while in the present Section $\partial_\nu$ with
$\nu = 1, 2$ stands for the derivatives with respect to the angles
only. To work out the Nambu--Goldstone modes explicitly we use
\ba
g^{-1} \partial_t g &=& -i \left( -\dot{\zeta} \sin \omega J_x
+ \dot{\omega} J_y + \dot{\zeta} \cos \omega J_z \right) \ , 
\nonumber \\
u^{-1} \partial_t u &=& -i \left(\dot{x} + {y \over 6}(x \dot{y}
-y \dot{x} ) \right) J_x \nonumber \\
&&-i \left(\dot{y} - {x \over 6}(x \dot{y} - y \dot{x}) \right) J_y
\nonumber \\
&&-i \left({1 \over 2}(y \dot{x}-x \dot{y}) \right) J_z \ ,
\label{F40}
\ea
and, from the Baker--Campbell--Haussdorf expansion,
\ba
U^{-1} \partial_\nu U &=& u^{-1} \partial_\nu u  \ , \nonumber \\
U^{-1} \partial_t U &=& u^{-1} \partial_t u + u^{-1} \left(g^{-1}
\partial_t g\right) u \nonumber \\
&=& -i\left(\dot{x} -\dot{\zeta} \sin \omega + {y\over 6}(x \dot{y} -
y \dot{x}) - y \dot{\zeta} \cos \omega \right) J_x \nonumber \\
&&-i \left(\dot{y} + \dot{\omega} - {x \over 6}(x \dot{y} - y \dot{x})
+ x \dot{\zeta} \cos \omega \right) J_y \nonumber \\
&&-i \left({1 \over 2}(y \dot{x} - x \dot{y})
+ \dot{\zeta} \cos \omega -x \dot{\omega} -y \dot{\zeta} \sin \omega
\right) J_z + \ldots \ .
\label{F41}
\ea
Here and in what follows, the dots indicate terms of higher order in
$\ve$. From Eqs.~(\ref{F40}) and (\ref{F41}) we obtain
\ba
a^x_t &=& \dot{x} +{y \over 6} (x \dot{y} - y \dot{x}) -\dot{\zeta}
\sin \omega - y \dot{\zeta} \cos \omega + \ldots\ , \nonumber \\
a^y_t &=& \dot{y} -{x \over 6} (x \dot{y} - y \dot{x}) + \dot{\omega}
+ x \dot{\zeta} \cos \omega + \ldots \ , \nonumber \\
a^z_t &=& -{1 \over 2} (x \dot{y} - y \dot{x}) + \dot{\zeta} \cos
\omega - y \dot{\zeta} \sin \omega - x \dot{\omega} + \ldots \ ,
\label{F42}
\ea
and 
\ba
a^x_\nu &=& \partial_\nu x +{y \over 6} (x \partial_\nu {y}
- y \partial_\nu {x})  +\ldots \ , \nonumber \\
a^y_\nu &=& \partial_\nu {y} -{x \over 6} (x \partial_\nu {y}
- y \partial_\nu {x}) +\ldots  \ , \nonumber \\
a^z_\nu &=& -{1 \over 2} (x \partial_\nu {y} - y \partial_\nu {x})
+ \ldots  \ . 
\label{F43}
\ea
Eqs.~(\ref{F42}) and (\ref{F43}) give the lowest-order contributions
to the Nambu-Goldstone modes for the parameterization~(\ref{F36}).

\subsection{Invariants}
\label{inv2}

As in Section~\ref{inv1} we build the effective Lagrangian upon
invariants constructed from the Nambu-Goldstone modes. To this end we
need to determine the behavior of these modes under transformations.
We consider a rotation $r$ about infinitesimal angles $\delta \chi_k$
around the space-fixed $k = x, y, z$ axes. We use Eq.~(\ref{F36}) for
$U$. With
\be
r g(\zeta, \omega) = g(\zeta', \omega') h(\gamma') \ , \ h(\gamma')
= \exp \{ i \gamma' J_z \} \ ,
\ee
\label{F44}
and $\zeta', \omega', \gamma'$ given below we have
\be
r U = r g(\zeta, \omega) u(x, y) = g(\zeta', \omega') h(\gamma') u =
g(\zeta', \omega') \ [ h(\gamma') u h^\dag(\gamma') ] \ h(\gamma') =
U' h \ .
\label{F45}
\ee
The last of equations~(\ref{F45}) defines $U' \equiv U(\zeta',\omega',
x', y')$ as an element of the coset space $SO(3) / SO(2)$. We
accordingly write
\ba
U'^{-1} i \partial_\mu U' &=& (a_\mu^x)' J_x + (a_\mu^y)' J_y +
(a_\mu^z)' J_z \ .
\label{F47}
\ea
Proceeding as in Section~\ref{inv1} we have
\ba
U^{-1} \partial_\mu U &\to& (r U)^{-1}\partial_\mu (r U) \nonumber \\
&=& (U' h)^{-1} \partial_\mu (U' h) \nonumber \\
&=& h^{- 1} (U')^{-1} (\partial_\mu U') h + h^{- 1} (\partial_\mu h) \ .
\label{F46}
\ea
From Eqs.~(\ref{F46}, \ref{F47}) and Eq.~(\ref{F39}) we obtain
\ba
\left(\begin{array}{c}
(a_\mu^x)' \\ (a_\mu^y)' \end{array} \right)
= \left( \begin{array}{cc}
\cos \gamma' & - \sin \gamma' \\
\sin \gamma' &  \cos \gamma' \end{array} \right)
\left( \begin{array}{c}
a_\mu^x \\ a_\mu^y \end{array} \right)
\label{F48}
\ea
and 
\be
(a_\mu^z)' = a_\mu^z + \delta_{\mu t} \dot{\gamma}' \ .
\label{F49}
\ee
We have used that $\gamma'$ as given in Eq.~(\ref{F52}) below is a
function of $t$ only.

It remains to work out the relation between the variables $\zeta',
\omega', x', y'$ and $\zeta, \omega, x, y$. A calculation similar to
that of Appendix~2 shows that $g(\zeta, \omega)$ transforms into
$g(\zeta', \omega')$ with $\zeta' = \zeta + \delta \zeta$ and $\omega'
= \omega + \delta \omega$ where
\be
\left(\begin{array}{c}
\delta \zeta \\ \delta \omega \end{array}\right)
= \left( \begin{array}{ccc} - \cot \omega \cos \zeta & - \cot \omega
\sin \zeta & 1 \\ - \sin \zeta & \cos \zeta & 0 \end{array}\right)
\left(\begin{array}{c}
\delta \chi_x \\ \delta \chi_y \\ \delta \chi_z \end{array}
\right) \ .
\label{F50}
\ee
According to Eq.~(\ref{F45}) the matrix $u$ transforms under the
action of $r$ into $h(\gamma') u h^\dag(\gamma')$. That transformation
differs from that of Eq.~(\ref{F7}) because of the prefactor $g$ in
the definition of $U$ in Eqs.~(\ref{F36}). Therefore, $x$ and $y$
transform into $x'$ and $y'$ according to
\be
\left(\begin{array}{c}
x' \\ y' \end{array}\right)
= \left( \begin{array}{ccc} \cos \gamma' & - \sin \gamma' \\ \sin
\gamma' & \cos \gamma' \end{array} \right)
\left(\begin{array}{c} x \\ y \end{array}\right)
\label{F51}
\ee
where
\be
\gamma' = {\cos \zeta \over \sin \omega} \delta \chi_x + {\sin \zeta
\over \sin \omega} \delta \chi_y \ .
\label{F52}
\ee
Eq.~(\ref{F51}) shows that $x^2 + y^2$ is invariant under rotations.
Moreover, since $\gamma'$ depends on time, under rotations the four
quantities $x, y, \dot{x}, \dot{y}$ are transformed into linear
combinations of $x', y', \dot{x}', \dot{y}'$.

We are now ready to construct the invariants. We begin with the time
derivatives in Eqs.~(\ref{F42}). Eq.~(\ref{F48}) shows that $(a^x_t)^2
+ (a^y_t)^2$ is invariant. We use the power counting of
Eqs.~(\ref{F38}) (see also Section~\ref{pow2} below) and drop terms of
order $\xi^2 \varepsilon$ and $\xi \Omega \varepsilon^k$ with $k \geq
3/2$. We also omit terms linear in $x, y, \dot{x}$, or $\dot{y}$ as
these vanish upon integration over $\theta$ and $\phi$, see
Eqs.~(\ref{F37}). That gives
\be
(a^x_t)^2 + (a^y_t)^2 \approx \dot{\omega}^2 + \dot{\zeta}^2 \sin^2
\omega + \dot{x}^2 + \dot{y}^2 + 2 (x \dot{y} - y \dot{x}) \dot{\zeta}
\cos \omega - {1 \over 3} (x \dot{y} - y \dot{x})^2 \ .
\label{F53}
\ee
The invariant form~(\ref{F53}) is the sum of three homogeneous
polynomials of orders zero, two, and four, respectively, in the
variables $x$, $y$, and their time derivatives. Under rotations, each
of these is transformed into another homogeneous polynomial of the
same order, see the text below Eq.~(\ref{F52})). Invariance of the
form~(\ref{F53}) implies that each of the said polynomials is
invariant by itself. Hence the invariants are
\ba
{\cal L}_{1a} &=& \dot{\omega}^2 + \dot{\zeta}^2 \sin^2 \omega \ ,
\nonumber \\
{\cal L}_{1b} &=& \dot{x}^2 + \dot{y}^2 + 2(x \dot{y} - y \dot{x})
\dot{\zeta} \cos \omega \ , \nonumber \\
{\cal L}_{1c} &=& (x \dot{y} - y \dot{x})^2 \ .
\label{F54}
\ea
The additional invariant
\be
{\cal L}_{1d} = (x^2 +y^2)[\dot{x}^2 + \dot{y}^2 + 2(x \dot{y} - y
\dot{x}) \dot{\zeta} \cos \omega]
\label{F55}
\ee
is obtained by multiplying ${\cal L}_{1b}$ with the invariant $(x^2 +
y^2)$. The invariant ${\cal L}_{1d}$ is of the same order as ${\cal
  L}_{1c}$.

We turn to the invariants constructed from the derivatives with
respect to the angles $\theta, \phi$ in Eqs.~(\ref{F43}). We confine
ourselves to terms of up to fourth order in $x$ and $y$ and their
derivatives. Eq.~(\ref{F48}) shows that for all $\nu = \theta, \phi$
the form $(a^x_\nu)^2 + (a^y_\nu)^2$ is invariant, and so are
$a^z_\nu$ and $a^z_\nu a^z_{\nu'}$. Forming suitable linear
combinations of these and multiplying with the additional invariant
$(x^2 + y^2)$ we find the invariants
\ba
{\cal L}_{2a} &=& (\vec{L} x)^2 + (\vec{L} y)^2 \ , \nonumber \\
{\cal L}_{2a'}&=& ({L_z} x)^2 + ({L_z} y)^2 \ , \nonumber \\
{\cal L}_{2b} &=& (x \vec{L} y - y \vec{L} x)^2 \ , \nonumber \\
{\cal L}_{2c} &=& (x^2 + y^2) \left( (\vec{L} x)^2 + (\vec{L} y)^2
\right) \ .
\label{F56} 
\ea
The construction of the invariants in Eqs.~(\ref{F54}) to (\ref{F56})
is based on the fact that under rotations, $u$ transforms into $h u
h^\dag$. This feature does not apply in the absence of any rotational
motion, i.e., for $g = 1$ or $\omega = 0 = \zeta$. It is easy to see
that in that case we would have $r U = r u = u(x', y') h$ where $x',
y'$ are nonlinear functions of $x, y$. The argument shows why the
invariants in Eqs.~(\ref{F54}) to (\ref{F56}) occur specifically in
the case of rotational motion but differ in cases like ferromagnetism
or paramagnetism where all modes considered are true Nambu-Goldstone
modes. It also shows that care is needed when considering the limit of
an infinitely large moment of inertia: In the framework of the present
formalism, that limit differs from the one where rotational motion is
ruled out from the outset.

In the construction of the invariants, time-reversal invariance has
been taken into account in the same manner as in Section~\ref{inv1}.
When we apply our formalism to atomic nuclei, an additional symmetry
(the ${\cal R}$-symmetry) comes into play. That symmetry matters for
the quantized version of the theory and is, therefore, deferred to
Section~\ref{disc}.

\subsection{Classical Field Theory}
\label{class2}

As in Section~\ref{class1} the effective Lagrangian $L$ is given in
terms of an arbitrary linear combination of the invariants constructed
in Section~\ref{inv2} and involves eight constants $C_i$, $i = a, b,
c, d$ and $D_i$, $i = a, a', b, c$ that must be determined by a fit to
data. In obvious notation we have
\ba
L &=& L_1 + L_2 = \int {\rm d} E \ {\cal L} \nonumber \\
&=& \int {\rm d} E \ \bigg( \sum_{i = a, b, c, d} \frac{C_i}{2} {\cal
L}_{1 i} - \sum_{i = a, a', b, c} \frac{D_i}{2} {\cal L}_{2 i} \bigg) \ .
\label{F57}
\ea
The integration over angles is defined in Eq.~(\ref{F21}).

In slight difference to Eqs.~(\ref{F27}) we expand the real variable
$x$ in two ways, either in spherical harmonics $Y_{L \mu} = (-)^\mu
Y^*_{L -\mu}$ or in terms of the real orthonormal functions
\ba
Z_{L \mu} \equiv \left\{\begin{array}{ll}
{1\over\sqrt{2}}\left(Y_{L \mu}+Y^*_{L \mu}\right) \ , & \mu>0 \ , \\
Y_{L 0} \ , & \mu=0 \ , \\
{1\over i \sqrt{2}}\left(Y_{L \mu}-Y^*_{L \mu}\right) \ , & \mu<0 \ .
\end{array}\right. 
\label{F58}
\ea
We note that the functions $Z_{L \mu}$ do not form the components of a
spherical tensor. We write
\ba
x &=& \sum_{L = 2}^\infty \sum_{\mu = -L}^L x_{L \mu} Z_{L \mu} \nonumber \\
&=& \sum_{L = 2}^\infty \sum_{\mu = -L}^L \tilde{x}_{L \mu} Y_{L \mu} \ ,
\label{F59}
\ea
and correspondingly for the real variables $y$, $\dot{x}$,
$\dot{y}$. As in Eqs.~(\ref{F27}), contributions with $L = 0$ and $L =
1$ are excluded. The coefficients $x_{l \mu}$ are real. For the
complex coefficients $\tilde{x}_{L \mu}$ we have $\tilde{x}^*_{L \mu}
= (-)^{\mu} \tilde{x}^{}_{L -\mu}$. For every value of $L$ the
coefficients $\tilde{x}_{L \mu}$ and $x_{L \mu'}$ are linearly related
in an obvious way. The first of Eqs.~(\ref{F59}) is useful for
quantization. The second is more useful when the calculation requires
angular-momentum algebra.

Using the first Eq.~(\ref{F59}) and carrying out the integration over
angles, we obtain for the total Lagrangian in Eq.~(\ref{F57})
\ba
L &=& {C_a\over 2} \left( \dot{\omega}^2 + \dot{\zeta}^2 \sin^2 \omega
\right) + {C_b\over 2} \sum_L \sum_\mu \left( \dot{x}_{L\mu}^2 +
\dot{y}_{L\mu}^2 \right) \nonumber \\
&+& C_b \dot{\zeta} \cos \omega \sum_L \sum_\mu \left(x_{L\mu}
\dot{y}_{L\mu} - y_{L\mu} \dot{x}_{L\mu} \right) \nonumber \\
&-& \frac{1}{2} \sum_{L \mu} \left( D_a L(L+1) + D_{a'} \mu^2 \right)
\left( x_{L \mu}^2 + y_{L \mu}^2 \right) \ .
\label{F60}
\ea
We have restricted ourselves to terms up to and including the orders
${\cal O}(\Omega)$, ${\cal O}(\varepsilon \Omega)$, and ${\cal
  O}(\xi)$. The canonical momenta are
\ba
p_\omega = \frac{\partial {\cal L}}{\partial \dot{\omega}} \ , \
p_\zeta = \frac{\partial {\cal L}}{\partial \dot{\zeta}} \ , \
p^x_{L \mu} = \frac{\partial {\cal L}}{\partial \dot{x}_{L \mu}}\ , \
p^y_{L \mu} = \frac{\partial {\cal L}}{\partial \dot{y}_{L \mu}} \ .
\label{F61}
\ea
In analogy to the first Eq.~(\ref{F59}) we define the real function
\ba
p_x(\theta, \phi) = \sum_{L = 2}^\infty \sum_{\mu = -L}^L p^x_{L \mu}
Z_{L \mu}(\theta, \phi) \ ,
\label{F62}
\ea
and correspondingly for $p_y(\theta, \phi)$.

We use the Noether theorem as in Section~\ref{class1} and find the
conserved quantities ($k = x, y, z$)
\ba
Q_k = \int {\rm d} E \ \left\{
{\delta \omega \over \delta \chi_k} p_\omega + 
{\delta \zeta \over \delta \chi_k} p_\zeta + 
{\delta x \over \delta \chi_k} p_x
+ {\delta y \over \delta \chi_k} p_y \right\} \ .
\label{F63}
\ea
We identify the $Q_k$ with the three components of angular momentum.
Explicitly we obtain from Eq.~(\ref{F50}) and from the differential
form of Eq.~(\ref{F51})
\ba
Q_x &=& - p_\omega \sin \zeta - p_\zeta \cot \omega \cos \zeta +
{\cos \zeta \over \sin \omega} \ K \ , \nonumber\\
Q_y &=& p_\omega \cos \zeta - p_\zeta \cot \omega \sin \zeta + {\sin
\zeta \over \sin \omega} \ K \ , \nonumber\\
Q_z &=& p_\zeta \ .
\label{F64}
\ea
Here
\ba
K = \int {\rm d} E \ (x p_y - y p_x ) \ .
\label{F65}
\ea
The terms in Eqs.~(\ref{F64}) that do not involve the factor $K$
correspond to the angular momentum of the rigid rotor. That is shown
below and is analogous to Section~\ref{prm}. The integral $K$ over
solid angle in Eq.~(\ref{F65}) is the angular momentum of the
two--dimensional oscillators that describe the surface vibrations.
That is easily shown by applying the Noether theorem to $\int {\rm d}
E \ (x^2 + y^2)$. As remarked below Eq.~(\ref{F52}), that expression
is invariant under $SO(2)$ transformations of $(x, y)$. The conserved
quantity associated with this invariance is $K$. The square of the
total angular momentum is
\be
Q^2 =p_\omega^2 + {1 \over \sin^2 \omega} \left( p_\zeta^2 - 2 K p_\zeta
\cos \omega + K^2 \right) \ .
\label{F66}
\ee

\subsection{Power Counting}
\label{pow2}

Given the full Lagrangian in Eq.~(\ref{F60}) we can now complete the
arguments leading to the relations~(\ref{F38}). To identify the terms
that are kept we use arguments similar to the ones in
Section~\ref{pow1}. In addition to Eqs.~(\ref{F38}) we assume
\ba
D_a, D_{a'}&\sim& \Omega/\varepsilon \ , \nonumber\\
C_a &\sim& \xi^{-1} \ , \nonumber \\
C_b, C_c, C_d &\sim& (\varepsilon \Omega)^{-1} \ ,
\label{F67}
\ea
which implies
\ba
C_a {\cal L}_{1a} &\sim& \xi \ , \nonumber \\
C_b {\cal L}_{1b} &\sim& \Omega \ , \nonumber \\
C_c {\cal L}_{1c}, C_d {\cal L}_{1d} &\sim& \varepsilon \Omega \ ,
\label{F68}
\ea
and
\ba
p_\omega, p_\zeta &\sim& 1 \ , \nonumber \\
p_x, p_y &\sim& \varepsilon^{-1/2} \ .
\label{F69}
\ea
If we were to scale $C_c \sim (\varepsilon^2 \Omega)^{-1}$ we would
find $C_c {\cal L}_{1c} \sim \Omega$, and that would be unexpected (or
``unnatural'') for a term with such a high power in the coordinates
$x, y$. We cannot completely rule out this possibility, however, as
effective field theories with unnaturally large scale do exist. As an
example we mention the pion-less nuclear effective field theory for
systems with a large scattering length~\cite{kaplanDB1997,kaplan1998}.
The large values of $p_x, p_y$ implied by the relations~(\ref{F69}) do
not seem natural but cannot be avoided if we insist on $x, y \sim
\varepsilon^{1/2}$.

To illuminate the role of the dimensionless parameter $\varepsilon$ we
consider the limit of an infinite system (with infinite moment of
inertia $C_a$, or $\xi \to 0$). The fields $\zeta$ and $\omega$
become static and have constant values that depend on the orientation
of the rotor. Spontaneous symmetry breaking gives rise to the
Nambu--Goldstone fields $x$ and $y$ that describe the low--energy
modes at the scale $\Omega$. The effective field theory breaks down at
the scale $\Lambda \gg \Omega$. The assumption $\varepsilon \sim
\Omega / \Lambda \ll 1$ takes account of this fact and is similarly
used in other non--relativistic applications of effective field
theory~\cite{roman1999, baer2004}. Terms of higher order in the time
derivatives of $x$ or $y$ that were suppressed in Eq.~(\ref{F53})
would be of order $\Omega \varepsilon^k$ with $k \geq 3/2$ and, thus,
of higher order in $\Omega / \Lambda$. A similar power counting for
the spatial derivatives results from the replacement $\vec{L} \to
\Omega \vec{L}$ as in Section~\ref{pow1}.

In the opposite case where $\Lambda\to \infty$ but where $\xi \ll
\Omega$ is finite (i.e., differs from zero), the terms of order
$\Omega \varepsilon \propto 1 / \Lambda$ disappear. 
The Hamiltonian describes a rotor coupled to a set of
oscillators. A problem occurs once the excitation energy is so large
that the amplitudes $x,y$ are of order unity and compete with the
finite rotations of the top. A distinction between the two types of
modes is then no longer meaningful, and spontaneous symmetry breaking
does not give an adequate description of the system.

\subsection{Effective Hamiltonian}
\label{eff2}

The effective Hamiltonian $H$ is obtained from the effective
Lagrangian $L$ via a Legendre transformation. To perform that
transformation we write the kinetic part $L_1$ of $L$ in a form that
displays its bilinear dependence on the velocities. We define the
infinite-dimensional velocity vector $V^T = \{ \dot{\zeta},
\dot{\omega}, \dot{x}_{L \mu}, \dot{y}_{L \mu} \}$ where $L = 2, 3,
\ldots$ and $\mu = L, L-1, \ldots, - L$ and write $L_1 = (1 / 2) V^T
\hat{G} V$. The matrix $\hat{G}$ is easily found from
Eq.~(\ref{F60}). We analogously define the vector of momenta $P^T = \{
p_\zeta, p_\omega, p^x_{L \mu}, p^y_{L \mu} \}$. Then the effective
classical Hamiltonian is
\ba
H = {1 \over 2} P^T \hat{G}^{-1} P + {1\over 2}\sum_{L \mu} \left( D_a
L(L+1) + D_{a'} \mu^2 \right) \left(x_{L\mu}^2 + y_{L\mu}^2\right) \ .
\label{F70}
\ea
To calculate the inverse $\hat{G}^{- 1}$ we write $\hat{G} = \hat{G}_0
+ \hat{G}_1$, where $\hat{G}_0$ is the diagonal part of $\hat{G}$, and
use perturbation theory in $\hat{G}_1$ so that $\hat{G}^{-1} =
\hat{G}_0^{-1} - \hat{G}_0^{-1} \hat{G}_1 \hat{G}_0^{-1} +
\hat{G}_0^{-1} \hat{G}_1 \hat{G}_0^{-1} \hat{G}_1 \hat{G}_0^{-1} \pm
\ldots$. Keeping only terms up to and including the orders
${\cal O}(\Omega)$, ${\cal O}(\varepsilon \Omega)$, and ${\cal
  O}(\xi)$ we obtain
\ba
H &=& {1 \over 2} \sum_{L \mu} \left[ {1\over C_b} \left((p^x_{L\mu})^2
+ (p^y_{L\mu})^2 \right) + \left(D_a L(L+1) + D_{a'} \mu^2 \right)
\left( x_{L\mu}^2 + y_{L\mu}^2 \right) \right] \nonumber\\
&& + {1 \over 2 C_a} \left[ p_\omega^2 + {1 \over \sin^2 \omega} \left(
p_\zeta^2 + 2 K p_\zeta \cos \omega + K^2 \cos^2 \omega \right) \right]
\nonumber \\
&=& {1 \over 2} \sum_{L\mu} \left[ {1\over C_b} \left( (p^x_{L\mu})^2 +
(p^y_{L\mu})^2 \right) + \left( D_a L(L+1) + D_{a'} \mu^2 \right)
\left( x_{L\mu}^2 + y_{L\mu}^2 \right) \right] \nonumber\\
&& + {1 \over 2 C_a} \left( Q^2 - K^2 \right) \ . 
\label{F71}
\ea
We have used Eq.~(\ref{F66}) and [see Eq.~(\ref{F65})]
\be
K = \sum_{L \mu} \left( x_{L \mu} p^y_{L \mu} - y_{L \mu} p^x_{L \mu}
\right) \ .
\ee
\label{F72}
In the Hamiltonian~(\ref{F71}), the Nambu-Goldstone modes undergo
harmonic vibrations. These are coupled via $K$ to a rigid rotor. The
vibrations are of order ${\cal O}(\Omega)$, while the rotations are of
order ${\cal O}(\xi)$.

We mention in passing that an alternative (and simpler) form of the
Legendre transformation seems to exist. Instead of defining the
momenta as in Eqs.~(\ref{F61}) and using Eq.~(\ref{F70}), one might
define the canonical momenta $p_x(\theta, \phi)$ and $p_y(\theta,
\phi)$ as functional derivatives of $\sum_i C_i{\cal L}_{1 i}$ with
respect to $x(\theta, \phi)$ and $y(\theta, \phi)$, use $(p_1, p_2,
p_3) = (p_\zeta, p_x, p_y)$ and $(q_1, q_2, q_3) = (\zeta, x, y)$, and
define $H$ as
\be
H = p_\omega \dot{\omega} + \int {\rm d} E \ \sum_{i = 1}^3 p_i
\dot{q}_i - L \ .
\label{F73}
\ee
That procedure yields the same result as the one used above only to
first order in $\hat{G}_1$. The terms of second order differ, and the
Hamiltonian resulting from Eq.~(\ref{F73}) is not rotationally
invariant. More precisely: Upon quantization $H$ does not commute with
the three components $Q_k$ of angular momentum in Eq.~(\ref{F64}).
Therefore, we have not used Eq.~(\ref{F73}).

\section{Quantized Hamiltonian}
\label{ham}

In the present Section we complete the program of the paper. We
quantize the effective Hamiltonian~(\ref{F71}), we discuss two
important discrete symmetries, and we investigate the resulting
spectra.

\subsection{Quantization}
\label{quant}
 
In quantizing $H$ and the three components $Q_k$ of angular momentum,
we encounter the problem that $\omega$ and $\zeta$ are curvilinear
coordinates, and that quantization in such coordinates is ambiguous,
see Ref.~\cite{kaplan1997} and references therein. The quantization
depends on the physical constraints that limit the dynamics to the
curved manifold, and thereby on the physical situation. We focus
attention on the relevant parts of $H$ and of the $Q_k$. These are
given by the Hamiltonian $H^{\rm rot}$ of the rigid rotor,
\ba
H^{\rm rot} = \frac{1}{2 C_a} \bigg( p_\omega^2 + \frac{1}{\sin^2
\omega} p^2_\zeta \bigg) \ ,
\label{F74}
\ea
and by the associated rigid-rotor parts of Eqs.~(\ref{F64}),
\ba
Q^{\rm rot}_x &=& - p_\omega \sin \zeta - p_\zeta \cot \omega \cos
\zeta \ , \nonumber\\
Q^{\rm rot}_y &=& p_\omega \cos \zeta - p_\zeta \cot \omega \sin
\zeta \ , \nonumber\\
Q^{\rm rot}_z &=& p_\zeta \ .
\label{F75}
\ea
We use the standard approach~\cite{papenbrock2011} to quantizing $H^{\rm rot}$.
In Appendix~4 we describe a different approach to quantization which
avoids the ambiguity encountered below and yields the same result.
With
\be
G^{-1} = \left( \matrix{ 1 & 0 \cr
                0 & 1 / \sin^2 \omega \cr} \right)
\label{F76}
\ee
we write $H^{\rm rot}$ in matrix form,
\be
H^{\rm rot} = \frac{1}{2 C_a} (p_\omega, p_\zeta) G^{-1} \left(
\begin{array}{c} p_\omega \\ p_\zeta \end{array} \right) \ .
\label{F77}
\ee
Quantization is achieved upon putting
\ba
\hat{H}^{\rm rot} &=& \frac{1}{2 C_a \sqrt{\det G}} (- i \partial_\omega,
- i \partial_\zeta) G^{-1} \sqrt{\det G} \left( \begin{array}{c} - i
\partial_\omega \\ - i \partial_\zeta \end{array} \right) \nonumber \\
&=& - \frac{1}{2 C_a} \bigg( \partial^2_\omega + \cot \omega
\partial_\omega + \frac{1}{\sin^2 \omega} \partial^2_\zeta \bigg) \ .
\label{F78}
\ea
The transition from Eq.~(\ref{F74}) to Eq.~(\ref{F78}) is tantamount to
putting
\ba
\label{F79}
\hat{p}_\omega &=& -i {1 \over \sqrt{\sin \omega}} \partial_\omega \sqrt{
\sin \omega} \ , \nonumber\\
\hat{p}_\zeta &=& -i\partial_\zeta \ .
\ea
The expressions~(\ref{F75}) must be symmetrized with respect to
$\zeta, p_\zeta$ prior to using Eqs.~(\ref{F79}). That gives
\ba
\hat{Q}^{\rm rot}_x &=& i \sin \zeta \, \partial_\omega +i \cot \omega
\cos \zeta \, \partial_\zeta \ , \nonumber\\
\hat{Q}^{\rm rot}_y &=& -i \cos \zeta \, \partial_\omega +i \cot \omega
\sin \zeta \, \partial_\zeta \ , \nonumber\\
\hat{Q}^{\rm rot}_z &=& -i\partial_\zeta \ .
\label{F80}
\ea
It is easy to check that the three components obey $[ \hat{Q}^{\rm
    rot}_x, \hat{Q}^{\rm rot}_y ] = i \hat{Q}^{\rm rot}_z$
(cyclic). Therefore, $\hat{\cal Q}^{\rm rot} = \{\hat{Q}^{\rm rot}_x,
\hat{Q}^{\rm rot}_y, \hat{Q}^{\rm rot}_z \}$ is a {\it bona fide}
angular-momentum operator.

For the remaining variables, we impose the usual quantization
conditions and choose a representation where all the $x$'s and $y$'s
are ordinary variables so that
\be
\hat{p}^x_{L \mu} = - i \frac{\partial}{\partial x_{L \mu}} \ , \
\hat{p}^y_{L \mu} = - i \frac{\partial}{\partial y_{L \mu}} \ .
\label{F81}
\ee
The components of the quantized angular momentum operator are given by
\ba
\hat{Q}_x &=& i \sin \zeta \, \partial_\omega +i \cot \omega \cos
\zeta \, \partial_\zeta + {\cos \zeta \over \sin \omega} \hat{K} \ ,
\nonumber\\
\hat{Q}_y &=& -i \cos \zeta \, \partial_\omega +i \cot \omega \sin
\zeta \, \partial_\zeta + {\sin \zeta \over \sin \omega} \hat{K} \ ,
\nonumber\\
\hat{Q}_z &=& -i\partial_\zeta \ ,
\label{F82}
\ea
with the operator $\hat{K}$ defined in Eq.~(\ref{F65}) and the
quantization condition~(\ref{F81}). The $\hat{Q}_k$ obey the
commutation relations
\be
[ \hat{Q}_x, \hat{Q}_y ] = i \hat{Q}_z \ (\rm cyclic) \ .
\label{F83}
\ee
The square of the total angular momentum operator is
\be
\hat{\cal Q}^2 = - \partial^2_\omega - \cot \omega \partial_\omega +
\frac{1}{\sin^2 \omega} ( - \partial^2_\zeta + 2 i \hat{K} \cos \omega
\partial_\zeta + \hat{K}^2) \ .
\label{F84}
\ee
It is obvious that $[\hat{Q}_k, \hat{K}] = 0$ for $k = x, y, z$. A
complete set of commuting angular-momentum operators is, thus,
$\hat{\cal Q}^2, \hat{Q}_z, \hat{K}$.  That is expected: For the
axially symmetric rotor the square $\hat{\cal Q}^2$ of the total
angular momentum and its projections $\hat{Q}_z$ onto the space--fixed
and $\hat{K}$ onto the symmetry axes are constants of the motion, with
quantum numbers $J (J + 1)$, $M$, $K$, respectively.

We expect that
\be
[\hat{Q}_k, \hat{H} ] = 0 \ {\rm for} \ k = x, y, z \ .
\label{F85}
\ee
To prove Eqs.~(\ref{F85}) we observe that the first line of
Eq.~(\ref{F71}) does not involve $\hat{p}_\omega$ or
$\hat{p}_\zeta$. Therefore, it suffices to show that the terms in this
line all commute with $\hat{K}$. That is straightforward. We conclude
that $\hat{H}$ consists of two commuting parts: The square of the
total angular momentum with quantum numbers $J(J + 1)$, $M$, $K$, and
the Hamiltonian for the surface vibrations which carry the quantum
number $K$. As a consequence, a rotational band occurs upon every
eigenstate of the vibrational part of $\hat{H}$. All rotational bands
have the same moment of inertia.  Differences arise only through terms
not considered in the approximation leading to Eq.~(\ref{F60}), see
Section~\ref{high}.

\subsection{Discrete symmetries}
\label{disc}

Discrete symmetries may restrict the spectrum beyond the requirement
imposed by time-reversal invariance. We follow Bohr and
Mottelson~\cite{bohr1975, bohr1982} and Weinberg~\cite{weinbergbook}.

We first consider ${\cal R}$-symmetry. That symmetry is realized if an
axially symmetric nucleus is, in addition, symmetric under a rotation
about $\pi$ around an axis perpendicular to the symmetry axis. For
definiteness, we choose a rotation $r$ around the $y$ axis. We
consider the product $g(\phi, \theta) r(0, \pi, 0)$ where (with
operators acting to the right) the rotation $r$ is applied prior to
$g$. We have
\ba
g(\phi,\theta) r(0,\pi,0) 
&=& r(\phi,\theta,0)g(0,\pi)\nonumber\\
&=& g(\phi+\pi,\pi-\theta)h(\pi) \ .
\label{F86}
\ea
We note that the naive evaluation $g(\phi, \theta) r(0, \pi, 0) =
g(\phi, \theta + \pi)$ would carry $\theta$ outside of its domain of
definition $0 \le \theta \le \pi$. Eqs.~(\ref{F86}) show that the
rotational degrees of freedom $\theta$ and $\phi$ behave under ${\cal
  R}$ as a particle on the sphere under the usual parity, i.e.
$(\theta, \phi) \to (\pi - \theta, \phi + \pi)$. The operation ${\cal
  R}$ acts also on the intrinsic variables $x, y$, and we have
\ba
u(x, y) r(0, \pi, 0) &=& u(-\kappa \sin \psi, \kappa \cos \psi)
r(0, \pi, 0) \nonumber \\
&=& g(\psi, \kappa) h(-\psi) r(0, \pi, 0) \nonumber \\
&=& g(\psi, \kappa) r(0, \pi, 0) h(\psi) \nonumber \\
&=& g(\psi + \pi, \pi - \kappa) h(\psi + \pi) \nonumber \\
&=& u(-(\pi - \kappa) \sin(\psi + \pi),(\pi - \kappa) \cos(\psi + \pi))
h(2 \psi) \nonumber \\
&=& u(-(\kappa - \pi) \sin \psi, (\kappa - \pi) \cos \psi) h(2 \psi)
\nonumber \\
&=& u(-x, -y) h(2 \psi) \ .
\label{F87}
\ea
Here, we used Eqs.~(\ref{C2}) that relate $g$ and $u$, and geometric
considerations in going from the second to the third line, and from
the sixth to the last line. Eqs.~(\ref{F87}) show that the intrinsic
variables transform under ${\cal R}$ as $(x, y) \to (-x, -y)$.
According to Bohr and Mottelson~\cite{bohr1975, bohr1982},
eigenfunctions of the intrinsic and external variables must have the
same ${\cal R}$ parity, i.e., both must simultaneously be either
positive or negative.

We next consider ordinary parity ${\cal P}$. Following Weinberg
(Ref.~\cite{weinbergbook}, Sect. 19.2), the fields $x$ and $y$
have the same parity as the generators $J_x$ and $J_y$. As components
of an axial vector, $J_x$ and $J_y$ have positive parity. Thus ${\cal
  P} (x, y) = (x, y)$. That implies that all quantized modes (i.e.,
single excitations, double excitations ...) of the fields $x$ and $y$
have positive parity and are allowed.

\subsection{Spectra}
\label{spec}

The quantized Hamiltonian is given by Eq.~(\ref{F71}) with the
understanding that the momenta are operators as defined in
Eqs.~(\ref{F79}) and (\ref{F81}). The term of leading order (${\cal
  O}(\omega)$) is given by the first line and written as
\ba
\hat{H}_\omega = \sum_{L \mu} \left( {(p^x_{L \mu})^2 + (p^y_{L \mu})^2
\over 2 C_b} + {C_b \over 2} \omega^2_{L \mu} \left( x_{L \mu}^2 +
y_{L \mu}^2 \right) \right) \ .
\label{F88}
\ea
It describes an infinite set of uncoupled harmonic oscillators with
frequencies $\omega_{L \mu} = [( L ( L + 1) D_a + \mu^2 D_{b}) /
  C_b]^{1/2}$. In practice, the breakdown scale $\Lambda$ serves as a
cutoff for this sum.  It is useful to combine $x_{L \mu}$ and $y_{L
  \mu}$ into a two-dimensional $SO(2)$-symmetric harmonic oscillator
with quantum numbers $n_{L \mu} = 0, 1, 2, \ldots$, $k_{L \mu} = 0,
\pm 1, \pm 2, \ldots$, and energies $(2n_{L \mu} + |k_{L \mu}| +1)
\omega_{L \mu}$. The intrinsic angular momentum of the oscillators is
given by the eigenvalues $K = \sum_{L \ge 2} \sum_{\mu = - L}^L k_{L
  \mu}$ of the operator $\hat{K}$, see Eq.~(\ref{F65}). The double sum
extends over occupied states only.

For the ground state all quantum numbers vanish. For the excited
states, we assume $D_b > 0$. The lowest vibrational state corresponds
to the single-quantum excitation of the mode $(x_{2 0}, y_{2 0})$. As
shown in Section~\ref{disc} the fields $x$ and $y$ have the same
positive parity as the corresponding generators $J_x$ and $J_y$ in
Eqs.~(\ref{F36}). Thus, the lowest vibrational state has $|K| = 1$ and
negative ${\cal R}$-parity. That is indeed observed in linear
molecules~\cite{herzberg1945}. Nuclei, however, are different. Here,
low-lying states are built from paired Fermions and have positive
${\cal R}$-parity. States with negative ${\cal R}$-parity correspond
to pair breaking and have high excitation energies at or above the
breakdown scale $\Lambda$ of the EFT. Thus, states with odd $K$ and
positive parity are absent in the low-energy spectrum of nuclei. As an
example we mention the absence of low-lying magnetic dipole
excitations~\cite{heyde2010, bentz2011}, i.e., $K = 1$ states with
positive parity, in the spectra of deformed even-even
nuclei~\cite{davidson1981, davidson1981b, aprahamian2006}.

For nuclei, two quanta need to be excited in the lowest $(x_{2 0},
y_{2 0})$ mode, yielding a degenerate pair of states with $K = 0$ and
$|K| = 2$. Data~\cite{davidson1981, aprahamian2006} indeed show that
the low-lying vibrations in even-even deformed nuclei have $K = 0$ and
$|K| = 2$. In the actinides and rare-earth nuclei, these states have
excitation energies of about 1~MeV, but are not degenerate, for two
reasons. First, anharmonicities, i.e. higher-order corrections to the
vibrational Hamiltonian~(\ref{F88}) lift any degeneracies, see
Section~\ref{high}. Second, the proximity of the breakdown scale at
$\Lambda \approx 2-3$~MeV amplifies such anharmonicities. Indeed,
there are only few nuclei that exhibit two-phonon excitations on top
of the one-phonon $K = 0$ or $|K| = 2$ vibrations, see
Refs.~\cite{sood1991, sood1992} for recent reviews. The impressive
spectra of $^{168}$Er~\cite{davidson1981} and of
$^{162}$Dy~\cite{aprahamian2006} confirm this picture. The
positive-parity states in those spectra must be viewed as
anharmonically distorted quantized vibrations corresponding to our
Nambu-Goldstone modes. These spectra also show negative-parity
states. These states cannot be understood within the EFT discussed in
this paper; they can possibly be viewed as vibrations on top of the
intrinsic odd-parity state with lowest energy.

The quantized version of the full Hamiltonian of Eq.~(\ref{F71}) is
\be
\hat{H}_{\omega,\xi}= \hat{H}_{\omega} + {\hat{I}^2 - \hat{K}^2
\over 2C_a} \ .
\label{F89}
\ee
The last term causes a rotational band to appear on top of each of the
vibrational states (band heads). The eigenfunctions are Wigner
$D$-functions $D_{M, K}^I(\alpha, \beta, 0)$ with total integer spin
$I$ and projections $-I \le M, K \le I$~\cite{varshalovich1988, papenbrock2011}. The
eigenvalues of $\hat{I}^2$ are $I (I + 1)$ with $I \ge |K|$. At this
order, all rotational bands have the same moment of inertia, and
deviations from this picture are due to higher-order
corrections, see Ref.~\cite{zhang2013} and Section~\ref{high}.

We conclude that the EFT predicts that in leading order, the
Nambu-Goldstone modes due to emergent breaking of rotational symmetry
yield a large number of harmonic vibrations. In next-to-leading order
each of these serves as head of a rotational band. All bands have
identical moments of inertia. Corrections of higher order considered
in Section~\ref{high} lead to anharmonicities of the vibrational
states and cause the moments of inertia to differ.

Given the close proximity of the breakdown scale in nuclei to the
vibrational excitation energy, it is reasonable to consider a simpler
-- but still model-independent -- approach to deformed nuclei. That
approach~\cite{papenbrock2011} uses an effective theory (as opposed to the
effective field theory of the present paper). It combines the
quantized rotations as lowest-energy excitations with the lowest
vibrational modes. The latter correspond to the $K = 0$ and $K = 2$
modes of the present paper. Thus, the effective theory replaces the
quantum fields $x$ and $y$ by their quantized modes of longest wave
length. Our results show that the effective theory is based upon a
solid field-theoretical foundation.

\subsection{Terms of Higher Order}
\label{high}

The extension of the effective field theory to higher-order terms is
straightforward but tedious. The program is this. (i) Use the power
counting of Section~\ref{pow2} to identify all terms that contribute
to the effective Lagrangian in a given order. That includes terms with
higher time derivatives. These can be treated by perturbative field
redefinitions~\cite{damour1991,grosseknetter1994}. (ii)
Expand the fields $x$ and $y$ into their normal modes as in
Eqs.~(\ref{F59}), and compute the Lagrangian by integration of the
Lagrangian density as in Eqs.~(\ref{F57}). The resulting expressions
are (complicated) sums involving Clebsch-Gordan coefficients. (iii)
Perform the Legendre transformation to the Hamiltonian within
perturbation theory to the desired order of the power counting. (iv)
Quantize the Hamiltonian and compute the spectrum.

Obviously, steps (i) to (iii) are quite laborious. Furthermore, the
kinetic part of the resulting Hamiltonian will have low-energy
constants that are complicated linear combinations of the
corresponding coefficients of the Lagrangian. The latter are not
known, and the former need to be determined from data. It is,
therefore, desirable to understand the transformation properties of
the coordinates $x_{L\mu}$ and $y_{L\mu}$ and of the canonical momenta
$p^x_{L\mu}$ and $p^y_{L\mu}$, and to directly construct the most
general Hamiltonian that is invariant under rotations at a given order
of the power counting.

The construction of the invariants is guided by the following
observations. Eqs.~(\ref{F51}) show that under rotations, the fields
$x(\theta, \phi)$ and $y(\theta, \phi)$ transform as the $x$ and $y$
components of a two-dimensional vector. These transformation
properties hold for every point $(\theta, \phi)$ on the unit
sphere. Using the expansion of the second of Eqs.~(\ref{F59}) we
conclude that the complex normal modes $\tilde{x}_{L \mu}$ and
$\tilde{y}_{L \mu}$ themselves, too, transform as the $x$ and $y$
components of a two-dimensional vector. The same is true for the
corresponding canonical momenta (denoted by $\tilde{p}^x_{L \mu}$ and
$\tilde{p}^y_{L \mu}$, respectively) as these stem from time
derivatives of the field modes $\tilde{x}_{L \mu}$ and $\tilde{y}_{L
  \mu}$. The integration $\int dE$ over the Lagrangian density in
Eq.~(\ref{F57}) singles out scalars. For instance, the invariant that
is bilinear in the momenta is
\ba
\sum_{L} \left( \tilde{p}^x_{L} \cdot \tilde{p}^x_{L} 
+ \tilde{p}^y_{L} \cdot \tilde{p}^y_{L} \right) &=& 
\sum_{L \mu} (-1)^\mu \left(\tilde{p}^x_{L \mu} \tilde{p}^x_{L -\mu}  
+ \tilde{p}^y_{L \mu} \tilde{p}^y_{L -\mu}  \right) \nonumber \\
&=& \sum_{L \mu} \left( ({p}^x_{L \mu})^2 + ({p}^y_{L \mu})^2 \right)
\ .
\label{F90}
\ea
Here $\tilde{p}^x_L$ and $\tilde{p}^y_L$ denote spherical tensors of
degree $L$ with components $\tilde{p}^x_{L \mu}$ and $\tilde{p}^y_{L
  \mu}$, respectively.

We apply these considerations first to the kinetic terms in
Eqs.~(\ref{F54}) and (\ref{F55}) and then to the potential terms in
Eqs.~(\ref{F56}). In calculating the kinetic part of the Hamiltonian
we encounter the need to invert the generalized form of the matrix
$\hat{G}$ in Eq.~(\ref{F70}). So far we have taken into account only
the terms ${\cal L}_{1 a}$ and ${\cal L}_{1 b}$ in
Eqs.~(\ref{F54}). As in Eqs.~(\ref{F57}) we now consider the sum
${\cal L}_1$ of all four terms in Eqs.~(\ref{F54}) and
(\ref{F55}). The kinetic part $L_1$ of the effective Lagrangian is
given by integration over ${\rm d} E$ of ${\cal L}_1$. We note that
${\cal L}_1$ is bilinear in the time derivatives of all dynamical
variables. We omit the term $\dot{\omega}^2$ which gives a trivial
contribution. Defining $\dot{x}_j$ as the totality of the time
derivatives $\{\dot{\zeta}, \dot{x}_{L \mu}, \dot{y}_{L \mu}\}$ we
proceed as in Section~\ref{eff2} and write
\be
L_1 = \frac{1}{2} \sum_{i j} \dot{x}_i A_{i j} \dot{x}_j \ . 
\label{F94}
\ee
In order not to overburden the notation we have chosen the letter $A$
rather than $G$ for the matrix defining $L_1$. As in
Section~\ref{eff2} we write $\hat{A}$ for the matrix and $A_{i j}$ for
its elements. We have $A_{i j} = A_{j i}$, and we write $\hat{A} =
\hat{A}^{(0)} + \hat{A}^{(1)}$. Here $\hat{A}^{(0)}$ ($\hat{A}^{(1)}$)
is the sum of the contributions arising from ${\cal L}_{1 a}$ and
${\cal L}_{1 b}$ (from ${\cal L}_{1 c}$ and ${\cal L}_{1 d}$,
respectively). The latter are small of order $\ve$ with respect to the
former. The momenta are defined by $p_i = \sum_j A_{i j} \dot{x}_j$,
and the kinetic part of the effective Hamiltonian is
\be
H_1 = \frac{1}{2} \sum_{i j} p_i (A^{- 1})_{i j} p_j \ . 
\label{F95}
\ee
We calculate $\hat{A}^{- 1}$ perturbatively and use the explicit form
of ${\cal L}_{1 c}$ and ${\cal L}_{1 d}$ in Eqs.~(\ref{F54}) and
(\ref{F55}). Details are given in Appendix~5. The contributions of
order $\Omega \ve$ to the effective Hamiltonian stemming from ${\cal
  L}_{1 c}$ (${\cal L}_{1 d}$) are denoted by $H_{1 c}$ ($H_{1 d}$,
respectively). We find
\ba
H_{1 c} &=& - \frac{C_{c}}{2 C^2_{b}} \int {\rm d} \Omega \ (x p_y -
y p_x)^2 \ , \nonumber \\
H_{1 d} &=& - \frac{C_{d}}{2 C^2_{b}} \int {\rm d} \Omega \ (x^2 + y^2)
(p^2_x + p^2_y) \ . 
\label{F96}
\ea
A contribution of order $\Omega \ve$ arises also from ${\cal L}_{1
  b}$. It is given by
\be
H_{1 b} = \frac{1}{2 C_a} \frac{\cos^2 \omega}{\sin^2 \omega} \int {\rm d}
\Omega \ (x p_y - y p_x)^2 \ .
\label{F97}
\ee

We turn to the potential terms. For the modes $x_{L \mu}$ and $y_{L
  \mu}$ we deal, in analogy to Eqs.~(\ref{F90}), with the invariant
$\sum_{L \mu} ( x^2_{L \mu} + y^2_{L \mu} )$. In evaluating the
remaining invariants in Eqs.~(\ref{F56}) we have to deal with the
angular-momentum operators $\vec{L}$ and $L_z$ acting on the fields
$x(\theta, \phi)$ and $y(\theta, \phi)$. Denoting by $L_\nu$ the
spherical components of $\vec{L}$, we have
\ba 
\hat{L}_\nu x(\theta, \phi) &=& \sum_{L \mu} \tilde{x}_{L \mu} \left(
\hat{L}_\nu Y_{L \mu}(\theta, \phi) \right) \nonumber \\
&=& \sum_{L \mu} \tilde{x}_{L \mu} C^{L \mu + \nu}_{1 \nu L \mu}
Y_{L \mu + \nu}(\theta, \phi) \ .  
\label{F91}
\ea 
Upon multiplication with $Y^*_{a \alpha}(\theta, \phi)$ (where $a$ and
$\alpha$ are arbitrary) and integration over ${\rm d} E$ we find
\ba
\int {\rm d}E \, Y^*_{a \alpha}(\theta, \phi) \hat{L}_\nu x(\theta, \phi)
&=& \sum_{L \mu} \tilde{x}_{L \mu} C^{L \mu + \nu}_{1 \nu L \mu} \int {\rm d} E
\, Y^*_{a \alpha}(\theta, \phi) Y_{L \mu + \nu}(\theta, \phi) \nonumber \\
&=& C^{a \alpha}_{1 \nu L \alpha - \nu} \tilde{x}_{a \alpha - \nu} \nonumber \\
&=& (-1)^\nu \hat{L}_{-\nu} \tilde{x}_{a \alpha} \ .
\label{F92}
\ea
In the last line the operator $\hat{L}_{-\nu}$ is understood to act on
the spherical tensor of rank $a$ with components $\tilde{x}_{a
  \alpha}$. Invariants built upon the normal modes $\tilde{x}_{L \mu}$
and $\tilde{y}_{L \mu}$ are, thus, obtained by viewing these modes as
components of spherical tensors of rank $L$. Scalars are constructed
after acting with $\vec{L}$ onto these components. For example
\ba
\int {\rm d}E \ (\vec{L} x)^2 &=& \sum_\nu (-1)^\nu \sum_{a \alpha}
\sum_{b \beta} \tilde{x}_{a \alpha} \tilde{x}_{b \beta} \int {\rm d}E \
C^{a \alpha - \nu}_{1 -\nu a \alpha} C^{b \beta + \nu}_{1 + \nu b \beta} 
Y_{a \alpha - \nu} Y_{b \beta + \nu} \nonumber \\
&=& \sum_\nu (-1)^\nu \sum_{a \alpha} \sum_{b \beta} \tilde{x}_{a \alpha}
\tilde{x}_{b \beta} (-1)^{\alpha - \nu} C^{a \alpha - \nu}_{1 -\nu a \alpha}
C^{b \beta + \nu}_{1 + \nu b \beta} \ \delta_a^b \
\delta_{-\alpha + \nu}^{\beta + \nu} \nonumber \\
&=& -\sum_{a \alpha} (-1)^\alpha \tilde{x}_{a \alpha} \tilde{x}_{a -\alpha}
\nonumber\\
&=& - \tilde{x}_L \cdot \tilde{x}_L \ .
\label{F93}
\ea
The evaluation of the terms in Eqs.~(\ref{F56}) is now
straightforward.  Upon applying the quantization rules, we obtain
higher-order terms in the Hamiltonian $\hat{H}$.

The terms in Eqs.~(\ref{F96}) and (\ref{F56}) do not depend on the
rotational degrees of freedom $\omega$ and $\zeta$. These terms lift
the degeneracies of the vibrational modes of excitation but do not
affect the moment of inertia. The term in Eq.~(\ref{F97}) depends
parametrically upon $\omega$ and couples the rotational bands with the
vibrational modes.

\section{Summary and Conclusions}

We have constructed the EFT for emergent symmetry breaking in deformed
nuclei. In addition to the Nambu-Goldstone modes, the theory contains
two additional degrees of freedom that describe rotations about the
body-fixed symmetry axis. Starting from a physically intuitive
and mathematically standard parameterization where
rotational and vibrational degrees of freedom are treated on an equal
footing, we have switched to a much more practical
parameterization where the rotational degrees of freedom receive a
separate treatment. The theory is characterized by three small
parameters.  These are (i) the ratio $\xi / \Omega$ of the energies of
rotational motion $\xi$ and of vibrational motion $\Omega$, (ii) the
ratio $\Omega / \Lambda$ where $\Lambda$ is the breakdown scale of the
EFT (typically given by the pairing energy or the energy of
single-particle excitation in the shell model), and (iii) the
parameter $\ve$ which characterizes deviations from harmonicity of the
vibrations. In lowest order, the spectrum consists of vibrations each
of which serves as head of a rotational band. The vibrations are due
to the quantized Nambu-Goldstone modes that describe the emergent
breaking of SO(3) symmetry. In leading order, the vibrational modes
are degenerate, and the rotational bands all have the same moment of
inertia. Terms of next order remove both degeneracies.

\ack This material is based upon work supported in part by the U.S.
Department of Energy, Office of Science, Office of Nuclear Physics,
under Award Numbers DE-FG02-96ER40963 (University of Tennessee), and
under contract number DEAC05-00OR22725 (Oak Ridge National
Laboratory).

\section*{Appendix 1: Equations of Motion in \\ Curvilinear
  Coordinates}

In curvilinear coordinates the equations of motion are obtained by
variation of the product of the Lagrangian density ${\cal L}$ and the
integration measure. We demonstrate that fact for the simplest
case. We consider a Lagrangian density ${\cal L}(\psi, \partial \psi /
\partial t, \partial \psi / \partial x_1, \ldots, \partial \psi /
\partial x_N)$ that depends on the field $\psi$ (a function of time
$t$ and of $N$ Cartesian variables $x_1, x_2, \ldots, x_N$), and on
the $N + 1$ derivatives of the field. Standard variation of the action
integral yields
\be
\int {\rm d} t \ \int \prod_{\nu = 1}^N {\rm d} x_\nu \ \bigg(
\frac{\partial}{\partial t} \frac{\partial {\cal L}}{\partial
(\partial_t \psi)} + \sum_{\mu = 1}^N \frac{\partial}{\partial x_\mu}
\frac{\partial {\cal L}}{\partial (\partial_{x_\mu} \psi)} +
\frac{\partial {\cal L}}{\partial \psi} \bigg) \delta \psi \ .
\label{A1}
\ee
We introduce $N$ curvilinear coordinates $\zeta_k(x_1, x_2, \ldots,
x_N)$ with $k = 1, \ldots, N$ that are functions of the $N$ Cartesian
coordinates $x_\mu$. We define the $N$--dimensional matrix
\be
M_{\mu k} = \frac{\partial x_\mu}{\partial \zeta_k} \ {\rm with} \ D
= \det M \ .
\label{A2}
\ee
Then
\be
\prod_{\nu = 1}^N {\rm d} x_\nu = D \prod_{k = 1}^N {\rm d} \zeta_k \ ,
\ \frac{\partial}{\partial x_\mu} = \sum_{k = 1}^N (M^{-1})_{k \mu}
\frac{\partial}{\partial \zeta_k} \ , \ \frac{\partial {\cal L}}
{\partial (\partial_{x_\mu})} = \sum_{l = 1}^N M_{\mu l} \frac{\partial
{\cal L}}{\partial (\partial_{\zeta_l})} \ .
\label{A3}
\ee
We insert all this into expression~(\ref{A1}) and obtain
\ba
&& \int {\rm d} t \ \int \prod_{k = 1}^N {\rm d} \zeta_k \ D \ \bigg(
\frac{\partial}{\partial t} \frac{\partial {\cal L}}{\partial
(\partial_t \psi)} \nonumber \\
&& \qquad + \sum_{\mu = 1}^N \sum_{l = 1}^N (M^{-1})_{l \mu}
\frac{\partial}{\partial \zeta_l} \sum_{n = 1}^N M_{\mu n}
\frac{\partial {\cal L}}{\partial (\partial_{\zeta_n} \psi)} +
\frac{\partial {\cal L}}{\partial \psi} \bigg) \delta \psi \ .
\label{A4}
\ea
The triple sum in expression~(\ref{A4}) can be written as
\be
\sum_{l = 1}^N \frac{\partial}{\partial \zeta_l}\frac{\partial
{\cal L}}{\partial (\partial_{\zeta_l} \psi)} + \sum_{\mu l n = 1}^N
(M^{-1})_{l \mu} \bigg\{ \frac{\partial}{\partial \zeta_l}  M_{\mu n}
\bigg\} \frac{\partial {\cal L}}{\partial (\partial_{\zeta_n} \psi)} \ . 
\label{A5}
\ee
The identities $ \det \ln D = \ln {\rm Trace} \ D$ and
$\partial_{\zeta_l} M_{\mu n} = \partial_{\zeta_n} M_{\mu l}$ imply
that expression~(\ref{A5}) is equal to
\be
\sum_{l = 1}^N \frac{\partial}{\partial \zeta_l}\frac{\partial
{\cal L}}{\partial (\partial_{\zeta_l} \psi)} + D^{-1} \bigg\{
\sum_{l =1}^N \frac{\partial}{\partial \zeta_l} D \bigg\}
\frac{\partial {\cal L}}{\partial (\partial_{\zeta_n} \psi)} \ .
\label{A6}
\ee
Using all that and the independence of $D$ of $t$ and $\psi$ we
rewrite expression~(\ref{A4}) as
\ba
\int {\rm d} t \ \int \prod_{k = 1}^N {\rm d} \zeta_k  \ \bigg(
\frac{\partial}{\partial t} \frac{\partial (D {\cal L})}{\partial
(\partial_t \psi)} + \sum_{l = 1}^N \frac{\partial}{\partial \zeta_l}
\frac{\partial (D {\cal L})}{\partial (\partial_{\zeta_l} \psi)} +
\frac{\partial (D {\cal L})}{\partial \psi} \bigg) \delta \psi \ .
\label{A7}
\ea
Comparing expression~(\ref{A7}) with expression~(\ref{A1}) we conclude
that variation of ${\cal L}$ in curvilinear coordinates is tantamount
to varying $D {\cal L}$ and otherwise treating the curvilinear
coordinates like Cartesian ones. That is what we use in
Section~\ref{class1}, with $D = \sin \theta$.

\section*{Appendix 2: The Matrix $M$}

We calculate the matrix $M$ defined by $\delta q_\nu = \sum_k M_{\nu
  k} \delta \chi_k$. We define
\ba
H &=& \omega \ [ \cos \zeta P_{x'} + \sin \zeta P_{y'}] \ , \nonumber \\
G &=& \omega \ [ - \sin \zeta P_{x'} + \cos \zeta P_{y'}] \ .
\label{B1}
\ea
Then
\be
[H, G] = - i \omega^2 P_{z'} \ , \ [H, -i P_{z'}] = G \ .
\label{B2}
\ee
With $r$ given by Eq.~(\ref{F23}) we write
\be
r \exp \{ - i H \} = \exp \{ - i \tilde{H} \} \exp \{ - i \delta \xi
\ P_{z'} \} \ .
\label{B3}
\ee
Here
\ba
\tilde{H} &=& \tilde{\omega} \ [ \cos \tilde{\zeta} \ P_{x'} + \sin
\tilde{\zeta} \ P_{y'} ] \ , \nonumber \\
\tilde{\omega} &=& \omega + \delta \omega \ , \nonumber \\
\tilde{\zeta} &=& \zeta + \delta \zeta \ .
\label{B4}
\ea
We calculate $\delta \omega, \delta \zeta$ and $\delta \xi$ to first
order in $\delta \chi_k$. Keeping only terms of first order we have
with $k = x', y', z'$
\be
\sum_k \delta \chi_k P_k = \frac{\delta \omega}{\omega} H + \delta
\xi \exp \{ - i H \} P_{z'} \exp \{ i H \} + \delta \zeta X
\label{B5}
\ee
where
\be
X = \bigg( \sum_{k = 0}^\infty \frac{(-i)^{k - 1}}{k!} \sum_{l = 0}^{k
- 1} H^l G H^{k - l - 1} \bigg) \exp \{ i H \} \ .
\label{B6}
\ee
Since $X$ must be of order zero in $H$ only terms with $l = k - 1$ in
Eq.~(\ref{B6}) contribute. From the commutation relations~(\ref{B2})
we have to zeroth order in $H$
\ba
H^l G &\to& G \omega^l \ {\rm for} \ l \ {\rm even} \ , \nonumber \\
H^l G &\to& - i P_{z'} \omega^{l + 1} \ {\rm for} \ l \ {\rm odd} \ . 
\label{B7}
\ea
Thus,
\be
X = ( - \sin \zeta P_{x'} + \cos \zeta P_{y'} ) \sin \omega + P_{z'} 
(\cos \omega - 1 ) \ .
\label{B8}
\ee
Similarly,
\ba
H^l ( - i P_{z'}) &\to& G \omega^{l - 1} \ {\rm for} \ l \ {\rm odd}
\ , \nonumber \\
H^l ( - i P_{z'}) &\to& - i P_{z'} \omega^l \ {\rm for} \ l \ {\rm
even} \ . 
\label{B9}
\ea
Therefore,
\be
\exp \{ - i H \} P_{z'} \exp \{ i H \} = P_{z'} \cos \omega + ( -
\sin \zeta P_{x'} + \cos \zeta P_{y'} ) \sin \omega \ .
\label{B10}
\ee
Inserting Eqs.~(\ref{B10}) and (\ref{B8}) into Eq.~(\ref{B5}) we find
\ba
\sum_k \delta \chi_k P_k &=& \delta \omega ( \cos \zeta P_{x'} +
\sin \zeta P_{y'}) - P_{z'} \delta \zeta \nonumber \\
&& + (\delta \xi + \delta \zeta) [ P_{z'} \cos \omega + ( - \sin
\zeta P_{x'} + \cos \zeta P_{y'} ) \sin \omega ] \ .
\label{B11}
\ea
We equate the coefficients multiplying $P_{x'}$, $P_{y'}$, and
$P_{z'}$ on both sides of that equation. With $\delta \omega = \delta
q_1, \delta \zeta = \delta q_2, \delta \xi + \delta \zeta = \delta
q_3$ we obtain
\be
\delta \chi_k = \sum_{l = 1}^3 (M^{-1})_{k l} \delta q_l
\label{B12}
\ee
where
\be
M^{-1} = \left( \matrix{
 \cos \zeta &  0                     & - \sin \zeta \sin \omega \cr
 \sin \zeta &  0                     &   \cos \zeta \sin \omega \cr
    0      &  - 1                   &   \cos \omega    \cr} \right) \ .
\label{B13}
\ee
The inverse matrix is
\be
M = \left( \matrix{
 \cos \zeta & \sin \zeta & 0 \cr
- \sin \zeta \cot \omega & \cos \zeta \cot \omega & - 1 \cr
- \frac{\sin \zeta}{\sin \omega} & \frac{\cos \zeta}{\sin \omega} & 0 \cr}
\right) \ .
\label{B14}
\ee

\section*{Appendix 3: Equivalence of the \\
Parameterizations~(\ref{F36}) and (\ref{F4}, \ref{F5}) for $U$}

We start from Eqs.~(\ref{F36}) and derive Eqs.~(\ref{F4}, \ref{F5}). We
define $\kappa \geq 0$ and $\Psi$ by
\ba
x &=& - \kappa \sin \Psi \ , \nonumber \\
y &=& \kappa \cos \Psi \ .
\label{C1}
\ea
We use a power-series expansion in $\kappa$, valid for $\kappa \ll 1$.
We prove the equivalence only to leading order in $\kappa$. Terms of
higher order can be treated analogously. In addition to
Eqs.~(\ref{F36}) we use the following definitions and identities,
valid for arbitrary values of $\kappa$, of the Euler angles $\alpha,
\beta, \gamma$, and of $\Psi$,
\ba
h(\gamma) &\equiv& e^{ i \gamma J_z} \ , \nonumber\\
r(\alpha, \beta, \gamma) &\equiv& g(\alpha, \beta) h(\gamma) \ ,
\nonumber \\
u(-\kappa\sin\Psi, \kappa\cos\Psi) &=& g(\Psi, \kappa) h^\dag(\Psi)
\ .
\label{C2}
\ea
The first of Eqs.~(\ref{C2}) defines $h(\gamma)$ for an arbitrary
angle $\gamma$. The second of Eqs.~(\ref{C2}) defines the rotation
$r(\alpha, \beta, \gamma)$ and shows that in Eqs.~(\ref{F36}), the
factor $g(\zeta, \omega)$ acts on $u$ like a rotation $r$ with third
Euler angle $\gamma = 0$. The third of Eqs.~(\ref{C2}) is an identity
for the function $u(x, y)$ defined in the third of Eqs.~(\ref{F36}).
That identity can easily be derived with the help of the relation
$\exp \{ - i \Psi \hat{J}_z \} \hat{J}_y \exp \{ i \Psi \hat{J}_z \} =
- \sin \Psi \hat{J}_x + \cos \Psi \hat{J}_y$.

We use the transformation law
\be
r(\zeta, \omega, 0) g(\Psi, \kappa) = g(\zeta', \omega') h(\gamma')
\ .
\label{C3}
\ee
Here $\zeta', \omega', \gamma'$ are functions of the angles $\zeta$
and $\omega$ and of the variables $\Psi$ and $\kappa$ and are given in
Ref.~\cite{papenbrock2011}. We have
\ba
\cot(\zeta' - \zeta) &=& \cos \omega \cot \Psi + \cot \kappa {\sin
\omega \over \sin \Psi} \ , \nonumber\\
\cos \omega' &=& \cos \kappa \cos \omega - \sin \kappa \sin \omega
\cos \Psi \ , \nonumber\\
\cot \gamma' &=& - \cos \kappa \cot \Psi - \cot \omega {\sin \kappa
\over \sin \Psi} \ .
\label{C4}
\ea
For $|\kappa| \ll 1$ that yields in leading order
\ba
\omega' &\approx& \omega + \kappa \cos \Psi \approx \omega + y \ ,
\nonumber \\ 
\zeta' &\approx& \zeta + {\kappa \sin \Psi \over \sin \omega} \approx
\zeta - {x \over \sin \omega} \ , \nonumber \\ 
- \gamma' &\approx& \Psi - \kappa \cot \omega \sin \Psi = \Psi + x \cot
\omega \ .
\label{C5}
\ea
Therefore,
\be
g(\zeta, \omega) g(\Psi, \kappa) h^\dag(\Psi) \approx g(\zeta - {x
\over \sin \omega}, \omega + y) h(x \cot \omega) \ .
\label{C6}
\ee
We use the third of Eqs.~(\ref{C2}) once again to write the result as
\ba
g(\zeta, \omega) g(\Psi, \kappa) h^\dag(\psi) &\approx& u \bigg(
-(\omega + y) \sin (\zeta - {x \over \sin \omega}), (\omega + y) \cos
(\zeta - {x \over \sin \omega}) \bigg) \nonumber \\
&& \qquad \times h(\Psi) h(x \cot \omega) \ .
\label{C7}
\ea
We recall that $U$ is defined in the coset space SO3/SO2. Hence
\be
\label{C8}
U \approx u \bigg( -(\omega + y) \sin (\zeta - {x \over \sin \omega}),
(\omega + y) \cos ( \zeta - {x \over \sin \omega})\bigg) \ .
\ee
Eq.~(\ref{C8}) gives the connection with the parameterization of $U$ in
Eqs~(\ref{F4}) and (\ref{F5}) to lowest order in $x, y$. Differences
in sign are due to the fact that here we work in the space-fixed
system. The variables $\omega_0$ and $\zeta_0$ in Eqs.~(\ref{F5}) are
seen to correspond to $\omega$ and $\zeta$, respectively. To lowest
order in $\kappa$, the variable $\omega_1$ corresponds to $y$ whereas
the variable $\zeta_1$ corresponds to $- x / \sin \omega$. The
occurrence of the factor $1 / \sin \omega$ in the last relation causes
the difficulties in the attempt to derive the equations of harmonic
motion directly from the parameterization~(\ref{F4}) via an expansion
in powers of $\omega_1$ and $\zeta_1$ and explains why we have
introduced the parametrization~(\ref{F36}). The calculation can
obviously be carried to higher orders. That establishes the complete
equivalence of the parameterizations of $U$ in Eqs.~(\ref{F36}) and
(\ref{F4}, \ref{F5}).

\section*{Appendix~4: Quantization in Curvilinear \\ Coordinates}

It is instructive to use another approach to quantization which shows
how the ambiguities that are associated with the
prescription~(\ref{F79}) are avoided. We assume that $\omega$ and
$\zeta$ are independent so that
\be
[ \omega, \zeta ] = 0 \ , \ [ \omega, p_\zeta ] = 0 \ , \
[ p_\omega, \zeta ] = 0 \ , \ [ p_\omega , p_\zeta ] = 0 \ . 
\label{D1}
\ee
For simplicity we have suppressed the symbol $\hat{}$ on the operators.
It remains to determine the commutators $[p_\omega, \omega]$ and
$[p_\zeta, \zeta]$. We choose a representation where $\omega$ and
$\zeta$ are ordinary real variables. Quantization is subject to three
requirements. (i) The expressions for $Q^{\rm rot}_k$ and for $H^{\rm
  rot}$ must be Hermitian. (ii) The $Q^{\rm rot}_k$ must obey the
standard commutation relations $[Q^{\rm rot}_x, Q^{\rm rot}_y] = i
Q^{\rm rot}_z$ (cyclic). (iii) When expressed in terms of the
quantized components $Q^{\rm rot}_k$ of angular momentum, the
Hamiltonian of the pure rotor must be given by Eq.~(\ref{F35}), with
$C \to C_a$. As for point (i), Hermitecity is defined with respect to
an integration measure for the variables $\omega$ and $\zeta$. The
matrix $U$ introduced in Eqs.~(\ref{F36}) is defined in the coset
space $SO(3) / SO(2)$. That suggests using for the volume element the
expression
\be
{\rm d} V = {\rm d} \omega \sin \omega \ {\rm d} \zeta \ . 
\label{D2}
\ee
An operator ${\cal O}$ is Hermitian if the equality
\be
\int {\rm d} V \ \Xi^{*} {\cal O} \Psi = \int {\rm d} V \
({\cal O} \Xi)^{*} \Psi
\label{D3}
\ee
holds for any two integrable functions $\Xi(\omega, \zeta)$ and
$\Psi(\omega, \zeta)$ that are periodic with period $2 \pi$ with
respect to both $\omega$ and $\zeta$. (That is why we have chosen in
Section~\ref{another} the ranges of integration as $0 \leq \omega_0,
\zeta_0 \leq 2 \pi$). Requirement (ii) on the commutators of the
$Q^{\rm rot}_k$ and the explicit form of the $Q^{\rm rot}_k$ in
Eqs.~(\ref{F75}) imply
\be
[ p_\omega, \omega ] = - i \ , \ [ p_\zeta, \zeta ] = - i \ .
\label{D4}
\ee
Eqs.~(\ref{F75}) show that for $Q^{\rm rot}_k$ to be Hermitian,
$p_\omega$ and $p_\zeta$ must be Hermitian, too. As formulated in
Eq.~(\ref{D3}), that condition is consistent with Eqs.~(\ref{D4}) if
$p_\omega$ and $p_\zeta$ obey Eqs.~(\ref{F79}). Calculating $H^{\rm
  rot}$ from Eq.~(\ref{F35}) with $C \to C_a$ we obtain the second of
Eqs.~(\ref{F78}). Thus, $H^{\rm rot}$ is the quantized Hamiltonian of
a rotor with $C_a$ the moment of inertia.

\section*{Appendix~5: Kinetic Part of the Effective Hamiltonian}

We use $\hat{A}^{- 1} \approx (\hat{A}^{(0)})^{- 1} -
(\hat{A}^{(0)})^{- 1} \hat{A}^{(1)} (\hat{A}^{(0)})^{- 1} +
\ldots$. With
\be
L_{1 b} = \sum_{L \mu} [(\dot{x}_{L \mu})^2 + (\dot{y}_{L \mu})^2] +
2 \dot{\zeta} \cos \omega \sum_{L \mu} [ x_{L \mu} \dot{y}_{L \mu} -
y_{L \mu} \dot{x}_{L \mu}]
\label{E1}
\ee
we have
\be
\hat{A}^{(0)} = \left( \matrix{ 
         \sin^2 \omega & - y_{L \mu} \cos \omega & x_{L \mu} \cos \omega \cr
         - y_{L \mu} \cos \omega &  1 & 0 \cr
         x_{L \mu} \cos \omega & 0 & 1 \cr} \right) \ .
\label{E2}
\ee
The non-diagonal terms $\propto x_{L \mu}, y_{L \mu}$ are of order
$\ve^{1 / 2}$. Therefore, we invert $\hat{A}^{(0)}$ by expanding in
powers of these terms and obtain to first order
\be
(\hat{A}^{(0)})^{- 1} \approx \left( \matrix{ 
         \frac{1}{\sin^2 \omega} & \frac{y_{L \mu} \cos \omega}{\sin^2
           \omega} & - \frac{x_{L \mu} \cos \omega}{\sin^2 \omega} \cr
         \frac{y_{L \mu} \cos \omega}{\sin^2 \omega} &  1 & 0 \cr
         - \frac{x_{L \mu} \cos \omega}{\sin^2 \omega} & 0 & 1 \cr} \right) \ .
\label{E3}
\ee
The matrix $\hat{A}^{(1)}$ receives contributions from both ${\cal
  L}_{1 c}$ and ${\cal L}_{1 d}$. The contribution from $L_{1 c}$ is
proportional to
\ba
&& \sum_{L L' L'' L''' L_1} \bigg[ \bigg( x_{L \mu} \dot{y}_{L' \mu'} -
\dot{x}_{L \mu} y_{L' \mu'} \bigg)^{L_1} \nonumber \\
&& \qquad \qquad \times \bigg( x_{L'' \mu''} \dot{y}_{L''' \mu'''} -
\dot{x}_{L'' \mu''} y_{L''' \mu'''} \bigg)^{L_1} \bigg]^0 \ .
\label{E4}
\ea
The upper indices on the big round and square brackets denote the
total angular momentum to which the terms are coupled. For $L_{1 d}$
we have correspondingly
\ba
&& \sum_{L L' L'' L''' L_1} \bigg[ \bigg( x_{L \mu} x_{L' \mu'} + y_{L \mu}
y_{L' \mu'} \bigg)^{L_1} \bigg( \dot{x}_{L'' \mu''} \dot{x}_{L''' \mu'''} +
\dot{y}_{L'' \mu''} \dot{y}_{L''' \mu'''} \bigg)^{L_1} \bigg]^0 \nonumber \\
&& + 2 \dot{\zeta} \cos \omega \sum_{L L' L'' L''' L_1} \bigg[ \bigg( x_{L \mu}
x_{L' \mu'} + y_{L \mu} y_{L' \mu'} \bigg)^{L_1} \nonumber \\
&& \qquad \qquad  \times \bigg( x_{L'' \mu''} \dot{y}_{L''' \mu'''} -
\dot{x}_{L'' \mu''} y_{L''' \mu'''} \bigg)^{L_1} \bigg]^0 \ .
\label{E5}
\ea
Both $L_{1 c}$ and $L_{1 d}$ are small. Therefore, we calculate $-
(\hat{A}^{(0)})^{- 1} \hat{A}^{(1)} (\hat{A}^{(0)})^{- 1}$ to lowest
order, i.e., by taking for $(\hat{A}^{(0)})^{- 1}$ only the diagonal
part of the matrix on the right-hand side of Eq.~(\ref{E3}). The
resulting terms in the effective Hamiltonian are then given by
\ba
H_{1 c} &=& - \frac{C_{c}}{2 C^2_{b}} \sum_{L L' L'' L''' L_1} \bigg[
\bigg( x_{L \mu} p^y_{L' \mu'} - p^x_{L \mu} y_{L' \mu'} \bigg)^{L_1}
\nonumber \\
&& \qquad \times \bigg( x_{L'' \mu''} p^y_{L''' \mu'''} - p^x_{L'' \mu''}
y_{L''' \mu'''} \bigg)^{L_1} \bigg]^0 \ , \nonumber \\
H_{1 d} &=& - \frac{C_{d}}{2 C^2_{b}} \sum_{L L' L'' L''' L_1} \bigg[
\bigg( x_{L \mu} x_{L' \mu'} + y_{L \mu} y_{L' \mu'} \bigg)^{L_1} \nonumber \\
&& \qquad \times \bigg( p^x_{L'' \mu''} p^x_{L''' \mu'''} + p^y_{L'' \mu''}
p^y_{L''' \mu'''} \bigg)^{L_1} \bigg]^0 \nonumber \\
&& - \frac{C_{d}}{2 C_{a} C_{b}} \frac{2 p_\zeta \cos \omega}
{\sin^2 \omega} \sum_{L L' L'' L''' L_1} \bigg[ \bigg( x_{L \mu}
x_{L' \mu'} + y_{L \mu} y_{L' \mu'} \bigg)^{L_1} \nonumber \\
&& \qquad \qquad  \times \bigg( x_{L'' \mu''} p^y_{L''' \mu'''} -
p^x_{L'' \mu''} y_{L''' \mu'''} \bigg)^{L_1} \bigg]^0 \ .
\label{E6}
\ea
The terms inversely proportional to $C^2_b$ are of order $\ve \Omega$,
the term inversely proportional to $C_a C_b$ is of order $\ve^2
\Omega$ and, therefore, negligible. That result justifies {\it a
  posteriori} the diagonal approximation for $(\hat{A}^{(0)})^{- 1}$.
With the help of the expansions~(\ref{F59}) and (\ref{F62}), we obtain
Eqs.~(\ref{F96}).

\section*{References}


\end{document}